# Estimation of the non-linear parameter in Generalised Diversity-Interactions models is unaffected by change in structure of the interaction terms

Rishabh Vishwakarma[1], Laura Byrne[1], John Connolly[2,3,4], Rafael de Andrade Moral[5] and Caroline Brophy[1]

[1]School of Computer Science and Statistics, Trinity College Dublin, Dublin 2, Ireland

[2]School of Mathematics and Statistics, University College Dublin, Dublin 4, Ireland

[3]German Centre for Integrative Biodiversity Research (iDiv) Halle-Jena-Leipzig, 04103 Leipzig, Germany

[4]Institute of Biology/Geobotany and Botanical Garden, Martin Luther University Halle-Wittenberg, Am Kirchtor

1, 06108 Halle (Saale), Germany

[5]Department of Mathematics and Statistics, Maynooth University, Maynooth, Co. Kildare, Ireland

**Corresponding Author**

Rishabh Vishwakarma

Email: vishwakr@tcd.ie

Address: School of Computer Science and Statistics, Trinity College Dublin, Ireland

## Abstract

Research over the past few decades has given rise to a broad consensus that there exists a strong association between the biodiversity in an ecosystem and its outputs (called ecosystem functions). The study of this association is known as the biodiversity and ecosystem function relationship. There are many approaches to modelling the biodiversity and ecosystem function relationship, for example, a linear modelling approach that assumes ecosystem function is driven by the number of species in the system. However, the definition of biodiversity can be broader than the number of species in the system and may include species abundances or species functional type distribution. Diversity-Interactions modelling is a regression-based framework that models the biodiversity and ecosystem function relationship by expressing ecosystem functions as a linear combination of species-specific effects, species' abundances, and species' interactions. The species interactions in a Diversity-Interactions model are expressed as being proportional to the product of the species proportions and can take several different forms ranging in complexity from a single interaction term (assuming all pairs of species interact in the same way) to many interaction terms (e.g. assuming a separate interaction for all pairs of species). The specification of the interactions may also include a non-linear parameter ($\theta$) as an exponent to the product of the species proportions to allow deviation from the assumption that the interactions must be directly proportional to the product of the species proportions, giving rise to Generalized Diversity-Interactions modelling. The structure of the interaction terms describes the underlying biological processes and thus the selection of a correct structure is important. In the absence of $\theta$ (i.e., when the $\theta$ exponent is equal to 1), this selection can be done using a series of hierarchical F-tests or other selection metrics, such as information criteria. The inclusion of $\theta$ introduces complexity to the model selection process: we could (a) select the interaction structure first by assuming $\theta = 1$ and then estimate $\theta$, or (b) estimate $\theta$ first and then select the appropriate interaction structure by fixing $\theta$, or (c) test for $\theta$ and its inclusion for each interaction structure. It is unknown whether the outcome of the model selection process is influenced by these various approaches. It is also unknown whether the estimation of $\theta$ is robust to changes in the structure of the linear interaction terms of the model. Using a simulation study, we test the robustness of $\theta$ and compare multiple model selection approaches to identify an optimal and computationally efficient model selection procedure for Generalized Diversity-Interactions models. Results show that estimation of $\theta$ is robust and remains unbiased when the underlying structure of interaction terms is changed and that the most efficient model selection procedure is to first estimate $\theta$ for one interaction structure and then reuse this estimate for the other interaction structures.

**Introduction:**

In recent decades, interest in quantifying the relationship between biodiversity and ecosystem function (BEF) has driven a wealth of experiments and associated statistical modelling approaches (Hector et al. 1999; Schmid et al. 2002; Bell et al. 2009; Kirwan et al. 2007; Cardinale et al. 2009; Isbell et al. 2011). Studies have shown that increasing the biodiversity of an ecosystem can improve the performance and stability of ecosystem functions across a range of ecosystem types (Tracy et al. 2004; Bell et al. 2005; Hooper et al. 2005; Balvanera et al. 2006; Worm et al. 2006; Cardinale et al. 2007; Finn et al. 2013; Gamfeldt et al. 2013). Species richness is often assumed to be the main driver of the BEF relationship (Spehn et al. 2005; Tillman et al. 1997b); however, community evenness, species' relative abundances, and their functional groupings may also be strongly influential (Wilsey and Polley 2004; Reich et al. 2004; Ebeling et al. 2014; Lembrechts et al. 2018). The Diversity-Interactions (DI; Kirwan et al. 2009; Brophy et al. 2011; Dooley et al. 2015; Brophy et al. 2017) and Generalized Diversity-Interactions approaches (GDI; Connolly et al. 2013) model the BEF relationship using a broader definition of species diversity by capturing species-specific effects, species' abundances, and species' interactions, in addition to richness patterns. The `DImodels` package (Moral et al. 2022) available for R software (R Core Team, 2021) can be used to fit and compare DI and GDI models. The species interactions in Diversity-Interactions models can take several different forms and may include a non-linear parameter. The structure of these interaction terms provides insight into the underlying biological processes and thus the proper estimation of the interaction structure is important. In this paper, we 1) explore and test (via simulation) the robustness of the non-linear parameter, and 2) compare different model fitting approaches for selecting the best interaction structure for Generalized Diversity-Interactions models in a computationally efficient way.

DI and GDI modelling are regression-based approaches that use species proportions and their interactions (defined as being proportional to the products of pairs of species proportions) as predictors to capture species divecrsity effects. Additional block or treatment effects may also be included. The specification of the interactions can range in complexity from a single interaction term (assuming all pairs of species interact in the same way) to many interaction terms (for example, each pair of species interacts uniquely) (Kirwan et al 2009) and may include a non-linear parameter if species interactions are not directly proportional to the product of their proportions, giving rise to Generalized Diversity-Interactions models (Connolly et al 2013). Figure 1 shows the effect of the non-linear parameter ($\theta$) on species interactions and model interpretation in GDI models using a hypothetical two-species example; the parameter $\theta$ moderates the realisation of the species interaction effects and can 'flatten' the interactions across the species proportions gradient. Model selection is an important part of modelling as it aids in a better understanding of the response-predictor relationship as well as the identification of significant and non-significant predictors (Mitchell and Beauchamp 1988). In the absence of the non-linear parameter ($\theta$ set equal to 1, not estimated), model selection for the interaction terms in DI models can be carried out through a series of hierarchical comparisons (Kirwan et al 2009). Further, there is a plethora of techniques available to perform model selection for linear regression; these include F-tests, AIC (Akaike 1973), BIC (Schwarz 1978), stepwise regression (Breaux 1967), etc. The inclusion of the non-linear term in GDI models complicates the model selection process: should the user first identify the most appropriate interaction and then estimate $\theta$, or should they first estimate $\theta$ and then select the interaction structure, or should $\theta$ and the interaction structure be estimated jointly. Estimating $\theta$ and the interaction structure jointly would be desirable, but is also computationally expensive. Hence for increased user-friendlyness we explore viability of the following three possible approaches to selecting the best model: (a) Select the appropriate interaction structure first by ignoring $\theta$ (i.e.,

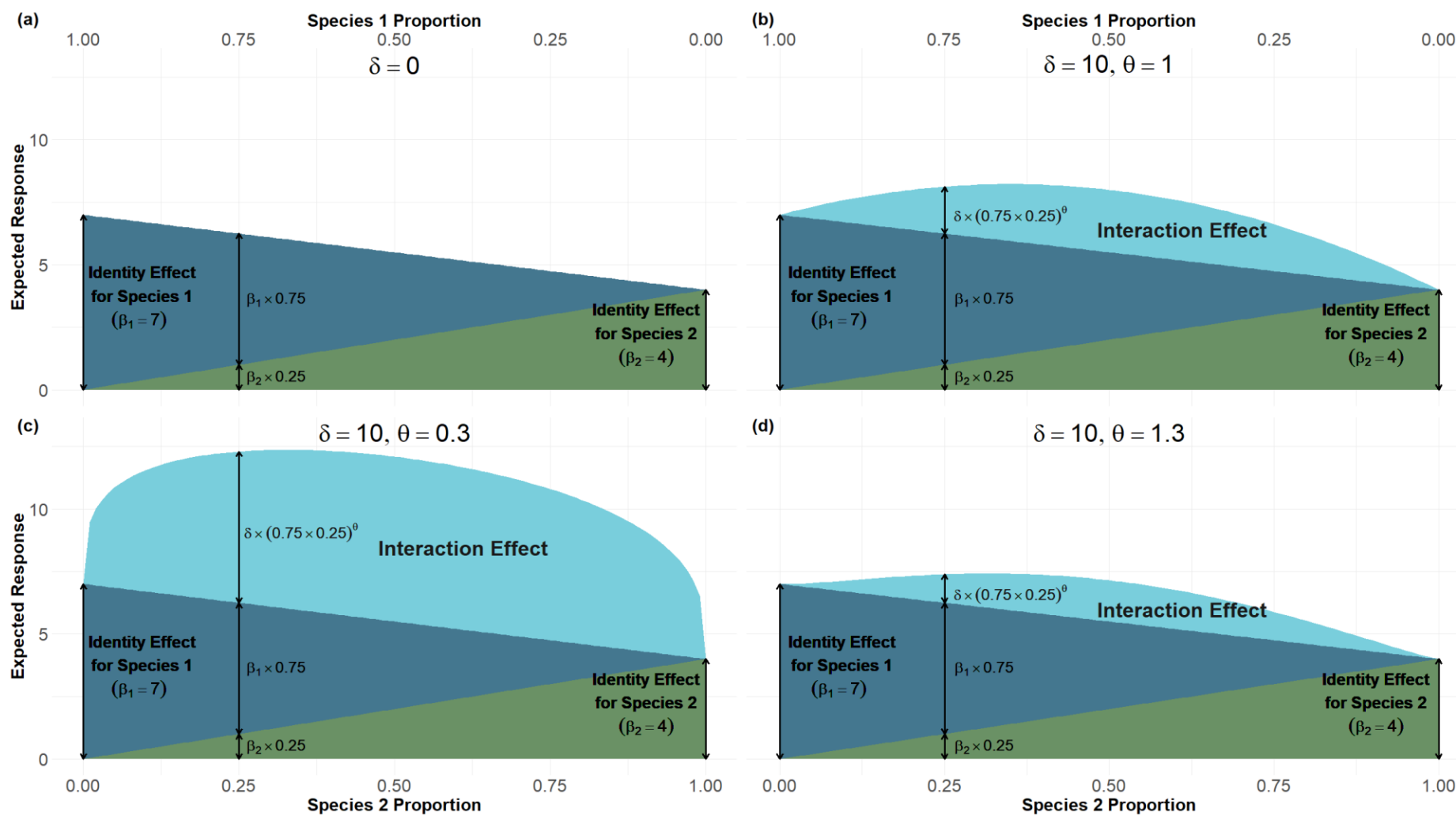

Figure 1 (Adapted from Grange et al., 2021). Illustration of the impact of the non-linear parameter (θ) on species interactions and interpretation in Generalized Diversity-Interactions models: A hypothetical two-species mixture is considered, with the response being yield, giving the equation $\hat{y} = \beta_1 P_1 + \beta_2 P_2 + \delta(P_1 P_2)^\theta$. The value of the response is expressed as a combination of the species identities and species interactions for all possible communities involving t two species, ranging from a monoculture of species 1 (on left) to a monoculture of species 2 (on right) and all possible two-

species mixtures in between. $\beta_1$ and $\beta_2$ are 'identity effects' for species 1 and species 2 respectively and are the expected performances of each species in monocultures. The expected performance of mixtures is the weighted average of the identity effects ($\beta_1 P_1 + \beta_2 P_2$) plus the interaction effect ($\delta * (P_1 P_2)^\theta$). The non-linear parameter ($\theta$) feeds into the interaction effect as an exponent which scales the product of the species proportions (($P_1 P_2)^\theta$). Four different scenarios for the expected response are presented: (a) No interaction term ($\delta = 0$, $\theta$ omitted), (b) No non-linear parameter ($\theta = 1$), (c) $\theta = 0.3$, and (d) $\theta = 1.3$; in each panel the identity effects are kept constant, in panels (b)-(d) the value of the interaction parameter ($\delta$) is kept constant (and a positive value used). The expected ecosystem function for an example 0.75:0.25 mixture is shown across the four scenarios and is computed as $\beta_1 \times 0.75 + \beta_2 \times 0.25 + \delta \times (0.75 \times 0.25)^\theta$. These concepts scale up for systems with more than two species.

assuming $\theta = 1$) and then estimate and test for the inclusion of $\theta$, (b) Estimate $\theta$ and test its inclusion for the simplest interaction structure first and then reuse that estimate to fit the remaining interaction structures and select the best model, and (c) Estimate $\theta$ and test its inclusion for each interaction structure and then perform model selection. Approach (c) is the most exhaustive method for model selection, but is computationally expensive, while approaches (a) and (b) aim for efficiency but rely on $\theta$ being invariant across varying specifications of species interactions making it important to test whether the estimate of $\theta$ is robust to changes in the structure of the interaction terms of the model. In this paper, we address the following two questions using a simulation study:

1) Is the estimation of the non-linear parameter ($\theta$) of a GDI model affected by changing the structure of the interactions?
2) What is the optimal and most computationally efficient model selection process for GDI models?

**Review of DI and GDI Models:**

The DI modelling framework (Kirwan et al. 2009) models the BEF relationship by expressing an ecosystem function response as a linear function of the relative abundances of the species spread across the simplex space (Kirwan et al. 2007, Cornell 2011). BEF data suitable for applying the DI models framework would include a range of experimental units (species communities) where species diversity is manipulated across dimensions such as species composition (identity), richness, and/or evenness to assess the impact of these variables on the ecosystem function. It is also possible to apply the DI modelling approach to appropriate observational data. The general formulation of a DI model is

$$y = \text{Identites} + \text{Interactions} + \text{Structures} + \varepsilon \qquad Eq\ 1$$

The response (y) is a community-level ecosystem function (e.g., biomass or weed resistance in a grassland ecosystem). The Identities and the Interactions components are the species-specific and the species interaction effects on the response, respectively, and are incorporated in the model using the initial proportions of the species and their products, respectively. Structures (experimental structures) are additional covariates or factors to capture experimentally manipulated treatments or blocks, or other measured descriptors of the experimental units. $\varepsilon$ is a normally distributed error term.

Connolly et al. 2013 showed that modifying the formulation of the species interaction terms in DI models leads to Generalised Diversity-Interactions (GDI) models that provide a more flexible framework for modelling BEF relationships. GDI models incorporate all the benefits of DI models and provide deeper insight into how individual pairs of species interact and by extension, affect community-level responses whilst also enabling us to explore phenomena such as the effects of diversity loss, functional stability, saturation properties of the BEF relationship, and transgressive overyielding (Connolly et al. 2013). DI models characterise the contribution of two species $i$ and $j$ to an ecosystem function as being proportional to the product of their relative abundances ($P_i P_j$), while GDI models assume a more general form for this contribution as $(P_i P_j)^\theta$, where $\theta$ is an additional parameter allowing for non-linearity in the relationship between the response and the interactions. A possible GDI model is:

$$y = \sum_{i=1}^{S} \beta_i P_i + \sum_{\substack{i,j=1 \\ i<j}}^{S} \delta_{ij} (P_i P_j)^\theta + \alpha A + \epsilon \qquad Eq\ 2$$

where $P_i$ is the sown proportion of species $i$, $s$ is the number of species in the system, $\beta_i$ is the identity effect of species $i$, the $\delta_{ij}$ parameters are the effects of the interactions between species $i$ and $j$, $A$ is a vector (or matrix) of experiment structures, $\alpha$ is a vector containing the effects of the experimental structures, and $\epsilon$ is a normally distributed error term with mean 0 and variance $\sigma^2$, i.e. $\varepsilon \sim N(0, \sigma^2)$. This variance is assumed to be constant, but it could be affected by the community structure; for example, it could differ for monoculture and mixture communities (Brophy et al. 2017; Cummins et al. 2021). $\theta$ is a non-linear parameter that can affect the nature of the relationship between the species interactions and the ecosystem function (Figure 1).

Equation 2 can be adjusted in multiple different ways by modifying the specification of the interaction terms to describe different biological hypotheses. These adjustments serve the purpose of reducing the number of interaction terms when the species pool is large. Table 1 gives a list of the several different GDI models (Kirwan et al. 2007, 2009; Connolly et al. 2013) along with their equations and the biological aspects that they describe. These models are a subset of a range of different possible models. Traditional model selection methods using F-tests or information criteria can be used to select the best model which strikes a balance between parsimony and explaining the BEF relationship. The models in Table 1 can be further expanded by crossing the identities and interaction terms with the variables such as year or treatment, as appropriate. They could also be extended to have multivariate responses (Dooley et al. 2015) or to include random pairwise interaction effects for modelling numerous species interactions (Brophy et al. 2017). Diversity-Interactions models have been widely used in understanding the BEF relationship in several experiments where the diversity was varied across the flora, fauna, or bacteria within the ecosystem (Kirwan et al. 2007; Connolly et al. 2009, 2011, 2018; Frankow-Lindberg et al. 2009; Nyfeler et al. 2009; O'Hea, Kirwan & Finn 2010; Brophy et al. 2011). A key advantage that DI and GDI models have over the other approaches, such as richness-only or anova models, is that they can be used to make predictions for the entire simplex space (provided the initial communities were sufficiently spread across the simplex space).

The $\theta$ parameter forms an integral part of GDI models. A value of $\theta = 1$ describes a linear interaction, proportional to the product of the species proportions, whilst a value of $\theta < 1$ corresponds to a stronger than expected contribution of species' pairs to ecosystem functioning, particularly at low abundances of the species, resulting in a stronger interaction effect. This is akin to a scenario where there is a strong niche separation of resources between the species resulting in little or no interspecific competition for the resources and is highlighted in Figures 1 and 2, which show the impact of varying the $\theta$ parameter in two-species and three-species systems respectively. Varying $\theta$ affects the 'flatness' of the interaction effects on ecosystem function across communities; e.g. in the middle column of ternary diagrams in Figure 2, for small $\theta$ values the species interaction effect is flatter for a larger range of communities across the entire simplex, in contrast to high $\theta$ values where the interaction effect is high for some communities in the centre and then declines as we move away from these central communities.

Table 1: Summary of Generalized Diversity-Interactions (GDI) models: Listed are some of the possible GDI Models, followed by their specific equations and the biological aspects they model. The non-linear parameter is $\theta$ and regulates the relationship between the species interactions and the ecosystem function. If $\theta = 1$, we get DI ([Kirwan et al. 2007, 2009](#)) models.

| Model | Equation |
|---|---|
| Null Model | $y = \beta + \alpha A + \epsilon$<br>No effect of changing diversity on ecosystem function. |
| Species Identity Model | $y = \sum_{i=1}^{S} \beta_i P_i + \alpha A + \epsilon$<br>Ecosystem function is affected only by species identities. |
| Average Pairwise Model | $y = \sum_{i=1}^{S} \beta_i P_i + \delta_{AV} \sum_{\substack{i,j=1 \\ i<j}}^{S} (P_i P_j)^\theta + \alpha A + \epsilon$<br>The strength of interactions is the same for all pairs of species and hence a single interaction term is sufficient. |
| Functional Group Model | $y = \sum_{i=1}^{S} \beta_i P_i + \sum_{q=1}^{T} \omega_{qq} \sum_{\substack{i,j \in FG_k \\ i<j}} (P_i P_j)^\theta + \sum_{\substack{q,r=1 \\ q<r}}^{T} \omega_{qr} \sum_{i \in FG_q} \sum_{j \in FG_r} (P_i P_j)^\theta + \alpha A + \epsilon$<br>Species interactions can be grouped based on the function they perform in the ecosystem. Assume T functional groups (FG$_1$ - FG$_T$), each with $n_t$ species, where $t = 1, \dots, T$. The parameter $\omega_{qq}$ is the interaction between two species from functional group $q$; and $\omega_{qr}$ is the interaction between two species from different functional groups, i.e., where $q \neq r$. |
| Additive Species Effects Model | $y = \sum_{i=1}^{S} \beta_i P_i + \sum_{\substack{i,j=1 \\ i<j}}^{S} (\lambda_i + \lambda_j)(P_i P_j)^\theta + \alpha A + \epsilon$<br>The contribution a species makes in its interaction is the same regardless of the species it interacts with. $\lambda_i$ is the contribution of species $i$ to the interaction with species $j$. |
| Full Pairwise Model | $y = \sum_{i=1}^{S} \beta_i P_i + \sum_{\substack{i,j=1 \\ i<j}}^{S} \delta_{ij} (P_i P_j)^\theta + \alpha A + \epsilon$<br>Each pair of species has a unique interaction effect on the ecosystem function. |

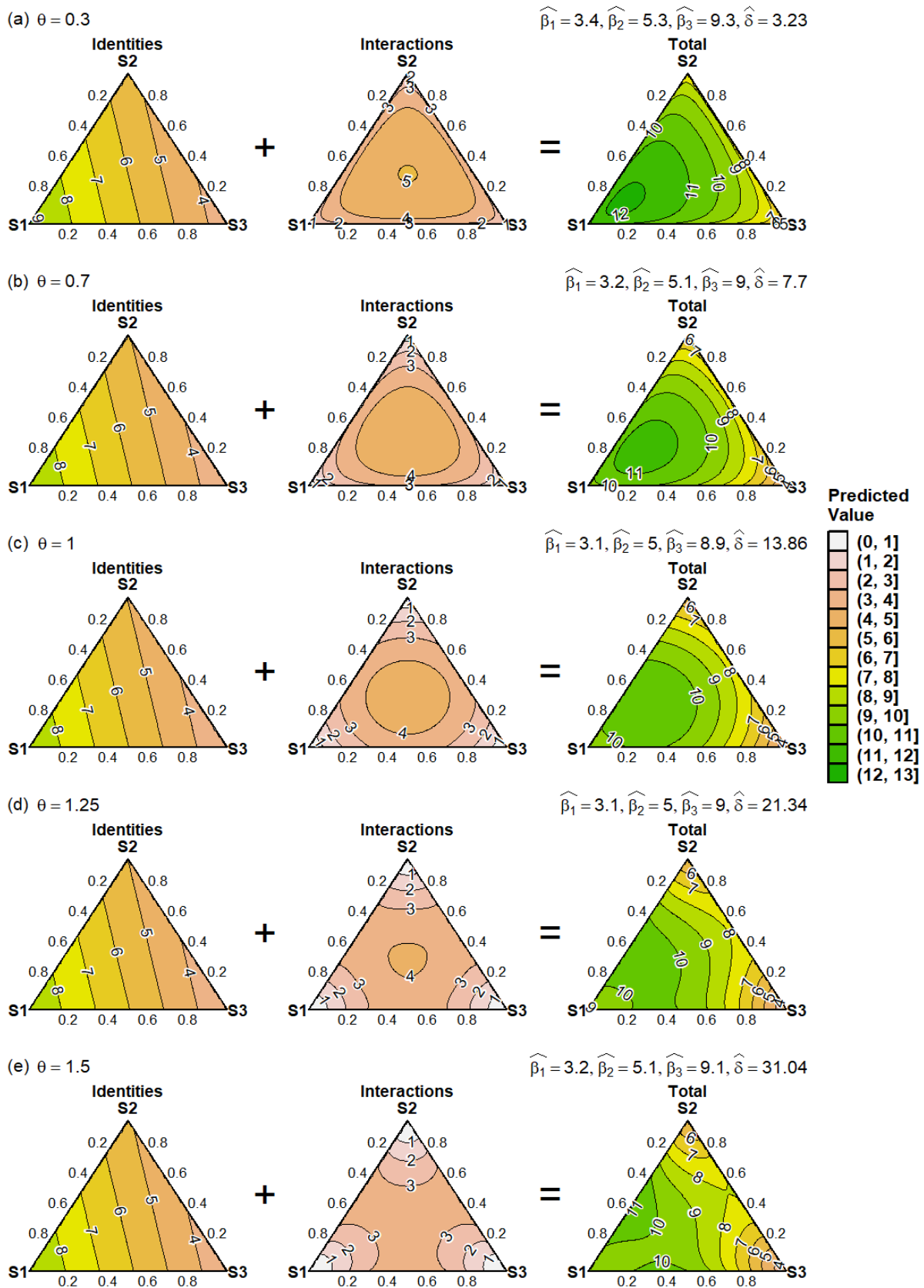

Figure 2. Ternary diagram illustrations of the effect of *θ* on an ecosystem function response in GDI models (final column) for a range of *θ* values and decomposed into the identity component

(first column) and interactions component (middle column). A single dataset was simulated from a three-species design assuming the average pairwise model, with identity effect coefficient values $\beta_1 = 9$, $\beta_2 = 5$, $\beta_3 = 3$, for species 1, 2 and 3 respectively, a value of 9 for the average interaction effect ($\delta_{AV}$), and $\theta$ equal to 0.8. The random error term added to the response from a normal distribution with $\mu = 0$ and $\sigma = 0.8$. Five versions of the average pairwise interaction model was then fit to this data, but where $\theta$ was not estimated, instead it was fixed for a range of values: (a) 0.3, (b) 0.7, (c) 1, (d) 1.15, and (e) 1.3 giving us five different estimated models (estimates of the $\beta_i$'s and $\delta_{AV}$ only and not the $\theta$ parameter). The model predictions across the simplex space are shown for each of these models, as well as the decomposition into identities and interactions components. The model parameter estimates are also shown for each row. The identity effects aren't strongly influenced by forcing the value of $\theta$ to change; however, the interaction component (and hence the total response) changes considerably depending on the forced value of $\theta$. For low values of $\theta$, the interactions (and hence total response) are high and are quite flat over a wide range of communities across the simplex.

**Methods:**

A simulation study was performed under two different experimental designs, one with four species and one with nine species. Under both designs, the true underlying model was assumed to be the full pairwise model with equation

$$y = \sum_{i=1}^{s} \beta_i P_i + \sum_{\substack{i,j=1 \\ i<j}}^{s} \delta_{ij}(P_i P_j)^\theta + \epsilon \qquad Eq\ 3$$

where $s = 4$ for the four-species design and $s = 9$ for the nine-species design, the $P_i$'s were the proportions of the respective species, $\beta_i$'s were the identity effects of the species, and $\delta_{ij}$ was the interaction effect between species $i$ and $j$, with $\theta$ and $\epsilon$ being the non-linear parameter and the random normal error term respectively.

For the four-species simulations, the species $SP_1$, $SP_2$, $SP_3$, and $SP_4$ were spread across the simplex space in a design that consisted of 37 different communities (shown in Figure 3). Each community was replicated three times and the design comprised of equi-proportional and imbalanced mixtures. The equi-proportional communities included the four monocultures, six two-species mixtures (50%, 50%, 0%, 0%), four three-species mixtures (33.33%, 33.33%, 33.33%, 0%), and one centroid community (25%, 25%, 25%, 25%). The imbalanced communities included four mixtures with each species being dominant in turn at three different levels of dominance (90%, 3.33%, 3.33%, 3.33%), (70%, 10%, 10%, 10%) and (40%, 20%, 20%, 20%), six mixtures with two species being dominant in turn at (40%, 40%, 20%, 20%), and four mixtures with three species being dominant in turn at (30%, 30%, 30%, 10%) giving a total of 22 imbalanced communities. The four species were assumed to be grouped into two functional groups (groupings based on the function they perform) with $SP_1$ and $SP_2$ being in the first functional group and $SP_3$ and $SP_4$ being in the second functional group.

For the nine species simulations, the species were named $SP_1$, $SP_2$, $SP_3$, …, $SP_9$ and the design opted for this model was the same as the one in the Jena dominance experiment (Roscher et al. 2005). There were nine monocultures, 36 two-species communities, 24 three-species communities, 18 four-species communities, 12 six-species communities, and one centroid community with nine species. This resulted in a simplex design with 100 unique equi-proportional communities. Each of these communities was then replicated thrice. The functional grouping structure assumed was $SP_1$, $SP_2$, $SP_3$, $SP_4$, and $SP_5$ being in functional group one, $SP_6$ and $SP_7$ being in functional group two and $SP_8$ and $SP_9$ being in functional group three. The full simplex design for both the four- and nine-species simulations are shown in Figures A1 and A2 in Appendix A.

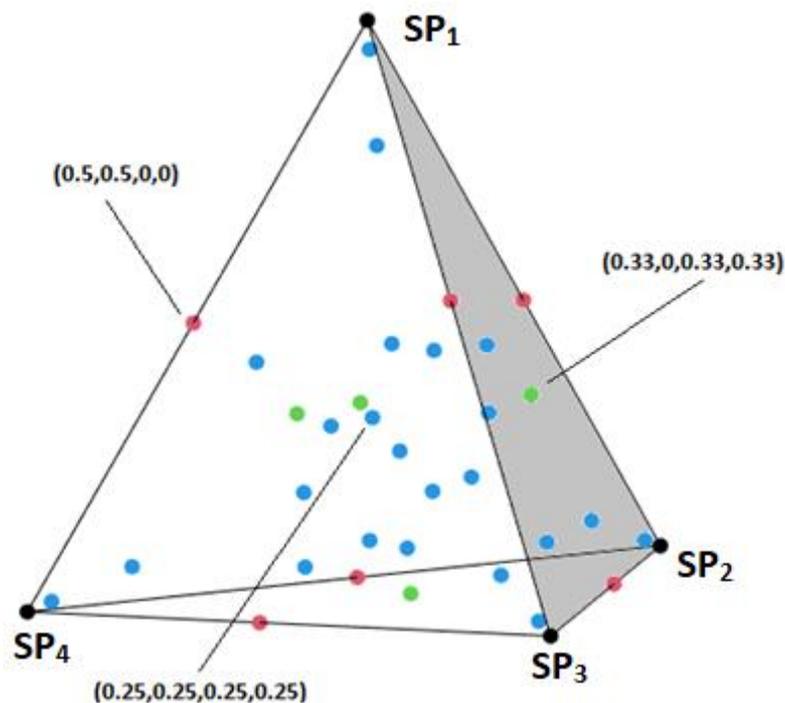

Figure 3. Graphical representation of the 4 species simplex design: Each point in the tetrahedron represents a four-species community and its position is determined by the relative abundances of the species (P1, P2, P3, and P4). The points are coloured according to the richness level of the community. The black points represent communities where richness=1, which are the monocultures and are positioned at the vertices of the tetrahedron. The red points represent the two-species mixtures and are positioned at points along the edges of the tetrahedron, determined by the relative abundances of the two species. The green points represent the three-species mixtures and are positioned along the faces of the tetrahedron. Finally, the blue points represent the four-species mixtures and are placed in the interior of the tetrahedron according to the relative proportions of the species.

The species identities and interaction effects used to simulate the responses for the four- and nine-species models are shown in Table 2a and Table 2b respectively. The identity effects for the four- and nine-species models were simulated from $N(3,9)$ and $N(7,4)$ distributions, respectively. To give a net non-zero interaction effect between the species, the interaction effects for the four-species model were simulated from a $N(8,16)$ distribution, while for the nine-species model they were simulated from a $N(9,36)$ distribution (the means and variances for these distributions were chosen to reflect values observed in real-world experiments). The functional groupings of the species weren't taken into consideration when simulating the interaction coefficients for these datasets as the true model was assumed to be the full pairwise model.

The response variable was simulated (using Eq 3 and the values in Table 2) for ten different $\theta$ values ranging from 0.05 up to 1.33. A random normal error term with mean 0 and constant standard deviation $\sigma$ was then added to the response. The value of $\sigma$ was varied to have five different values: 0.8, 0.9, 1.0, 1.1 and 1.2. This gave us a total of 100 settings for the full simulation study (2 experimental designs x 10 $\theta$ values x 5 $\sigma$ values). A total of 200 datasets were simulated for each setting. These datasets were simulated in R (version 4.0.3) and reproducible scripts are available in Appendix D and at https://github.com/rishvish/Theta-Simulation-Study. Increasing the number of simulated datasets up to 1000 was tested for a small number of settings, but the results stabilised at around 200 simulations and hence 200 was chosen as the number of datasets to simulate per setting.

Table 2: Coefficients of the four- (a) and nine- (b) species models (Eq 3). The shaded cells along the diagonal represent the identity effects. The value in cell $(i,i)$ corresponds to the coefficient of the identity effect of species $SP_i$. The value in cell row $i$ and column $j$, i.e. $(i,j)$, represents the coefficient of the interaction effect between species $SP_i$ and $SP_j$.

| (a) | SP₁ | SP₂ | SP₃ | SP₄ |
|---|---|---|---|---|
| SP₁ | 5 | 4.68 | 13.74 | 8.52 |
| SP₂ |  | 7 | 3.89 | 16.22 |
| SP₃ |  |  | 6 | 10.32 |
| SP₄ |  |  |  | 3 |

| (b) | SP₁ | SP₂ | SP₃ | SP₄ | SP₅ | SP₆ | SP₇ | SP₈ | SP₉ |
|---|---|---|---|---|---|---|---|---|---|
| SP₁ | 7 | 11.99 | 7.64 | 8.42 | -2.6 | 7.89 | 8.32 | 10.25 | 4.06 |
| SP₂ |  | 8 | 4.18 | -1.03 | 10.08 | 5.22 | 13.04 | 8.19 | 13.42 |
| SP₃ |  |  | 6 | 5.13 | 18.75 | 12.41 | 8.86 | 19.61 | 1.64 |
| SP₄ |  |  |  | 9 | 5.98 | 9.67 | 11.22 | 18.76 | 4.6 |
| SP₅ |  |  |  |  | 5 | 9 | 9.85 | 7.37 | 12.59 |
| SP₆ |  |  |  |  |  | 6 | 14.59 | 14.98 | 9.58 |
| SP₇ |  |  |  |  |  |  | 6 | -1.57 | 8.21 |
| SP₈ |  |  |  |  |  |  |  | 6 | 8.8 |
| SP₉ |  |  |  |  |  |  |  |  | 7 |

To test the robustness of $\theta$ estimation across different interaction specifications, the final four GDI models in Table 1 (the average pairwise, functional group, additive species and full pairwise models) were fit to each of the 200 simulated datasets for each 100 simulation settings using the DI function from the `DImodels` package (v1.2; Moral et al. 2022) in R (R Core Team, 2021). The value of $\theta$ was estimated by maximising the profile log-likelihood using the `DImodels` package and the distributional properties of the estimator were assessed graphically to determine whether the estimate of $\theta$ differed across the different models and from its true underlying value. Profile log-likelihood confidence intervals (CI) were also calculated. This is the interval for $\theta$ where the log-likelihood function $l(\theta)$ is greater than $l_{max}(\theta) - 0.5 \times \chi^2_{1-\alpha}(1)$. Here, $l_{max}(\theta)$ is the maximum log-likelihood value and $\chi^2_{1-\alpha}(1)$ is the $(1-\alpha) \times 100\%$ percentile of the chi-squared distribution with 1

d.f. The corresponding coverage of the CI was assessed by taking the proportion of times that the true value of $\theta$ fell within the computed CI.

To explore the efficacy of different model selection procedures, using the same 200 simulated datasets for each of the 100 simulation settings, we checked the proportion of times that the true underlying interaction structure was selected as the best model using three different model selection procedures. Table 3 gives a detailed description of these model selection procedures.

Table 3: The three model selection procedures tested for selecting the best interaction structure for GDI models along with their algorithmic description.

| Procedure | Description |
|---|---|
| (a) | (i) Assume $\theta = 1$ and fit all interaction structures first<br>(ii) Select the most appropriate interaction structure using AIC<br>(iii) Estimate $\theta$ for selected interaction structure<br>(iv) Test whether it significantly differed from 1 |
| (b) | (i) Estimate $\theta$ first for the simplest interaction structure<br>(ii) Test whether this estimate significantly differs from 1<br>(iii) If $\theta$ was significantly different from 1 reuse estimate of $\theta$ from step (i), else assume $\theta = 1$<br>(iv) Use value of $\theta$ deduced from step (iii) to fit the remaining interaction structures<br>(v) Use AIC to select the best interaction structure |
| (c) | (i) Estimate $\theta$ for each interaction structure<br>(ii) Test whether the $\theta$ estimates significantly differ from 1 for each interaction structure<br>(iii) Use AIC to select the best interaction structure |

Additional simulations for both the robustness of $\theta$ estimation and the model selection efficacy were carried out under different conditions, including a high number of species (up to 72), the presence of experimental structures, different true underlying models to the full pairwise interaction model, different structure of functional groupings, and higher variance of error terms. Simulations were also performed to test the robustness of a re-parameterisation of $\theta$ suggested by Connolly et al. 2018, where the $\delta$ coefficients are scaled by a factor of $\frac{2s^{2\theta}}{s(s-1)}$ to reduce the correlation between $\theta$ and $\delta$ coefficients.

**Results:**

For each of the four models (average pairwise interaction, functional group effects, additive species contributions, and full pairwise interactions) fit to the datasets in the simulation study, the mean estimate of $\theta$ was almost identical and was approximately equal to the true value of $\theta$ (Figure 4 and Table 4). Splitting the results up by the five $\sigma$ values (0.8, 0.9, 1, 1.1, and 1.2), it was found that the results were invariant to a changing $\sigma$, with the only effect of $\sigma$ being an increase in the variation of the distribution of the estimates of $\theta$ as the value of $\sigma$ increased (select results shown in Figure A3 in Appendix A). The results obtained from the study were similar for both the four- and nine-species cases. The mean estimate of $\theta$ was approximately equal to the true value of $\theta$ and the average coverage of the 95% confidence interval for the estimated $\theta$ was close to 1 for low values of $\theta$ and approached 0.95 as the value of $\theta$ increased, for both the four- and nine-species cases (Table 4). The unusual coverages for low $\theta$ values were due to a combination of convergence problems near the boundary of $\theta = 0$ and the interval being too precise (see Appendix A for more details). The standard deviations of the estimates for $\theta$ tend to increase as the true value of $\theta$ increases. This is because we are simulating different datasets for each value of $\theta$ and the range of the response variable is different for each value of $\theta$, which causes the change in the standard deviations of the estimates. Scaling the standard deviations of predicted estimates by the interquartile range for each unique $\theta$-model combination results in the standard deviations being similar for each unique $\theta$-model combination (Figure A5 in Appendix A).

The simulations for testing the efficacy of different model selection methods showed that method 'b', where we first estimate $\theta$ and then select the best interaction structure, was better than method 'a', where we select the appropriate interaction structures first and then estimate the value of $\theta$. For model selection procedure 'a', we found that for lower values of $\theta$ ($\theta < 0.5$), irrespective of the number of species and the underlying true structure of interaction terms, the average pairwise interaction model was selected as the chosen model almost 100% of the time. However, as the value of $\theta$ increased, the proportion of times that the true underlying (full pairwise in our example) interaction structure was chosen increased (Figure 5). Different selection metrics besides AIC, like F-tests and BIC, were also tested, but similar results were observed. A possible reason for this could be that for low values of $\theta$ ($\theta < 0.5$), the initial assumption of $\theta$ being equal to 1 is incorrect and thus all estimated models fit the data poorly and the selection criteria ends up selecting the model with the simplest structure, resulting in the average pairwise interaction model being selected every time. As the value of $\theta$ increases over 0.5, the initial assumption of $\theta$ being 1 isn't far off from the true value of $\theta$ and thus the models fit the data better and the selection metrics have more power to select the best interaction structure.

Model selection procedures 'b' and 'c' offered an improvement on this as instead of assuming $\theta$ to be 1, we first estimate it for a specific interaction structure and then reuse that estimate of $\theta$ to fit the remaining interaction structures in method 'b' or estimate $\theta$ separately for each interaction structure in method 'c'. Thus, these model selection procedures outperformed method 'a' and selected the true underlying interaction structure most of the time across the different values of $\theta$ (Figure 6 for method 'b' and Figure 7 for method 'c'). The proportion of times the true interaction structure gets selected did decrease sometimes as the value of $\sigma$ increased and this can be expected as increasing $\sigma$ had the effect of increasing the noise in the data. Similar results were observed when testing with different selection criteria besides AIC. Comparing the efficacy that model selection procedures 'b' and 'c' showed that method 'b' had comparable performance to method 'c' whilst giving a four- to seven-fold (depending on the number of species in the experiment) reduction in computation time (Figure A5 in appendix A). Thus, our recommendation is to use approach 'b' for model selection within the Diversity-Interactions modelling framework.

Simulations testing additional scenarios like a higher number of species (up to 72 species), a different true underlying interaction structure, a different structure of functional groups, the presence of additional experimental structures, and the reparameterisation of $\theta$, all yielded similar results in that the mean estimate of $\theta$ was approximately equal to the true value of $\theta$ and didn't differ much across the four (average pairwise interactions, functional group effects, additive species contributions and full pairwise interactions) estimated models. Results for model selection too were similar to the results observed for the four- and nine-species cases (Tables and figures presented in Appendix B for $\theta$ reparameterization and Appendix C for all other factors).

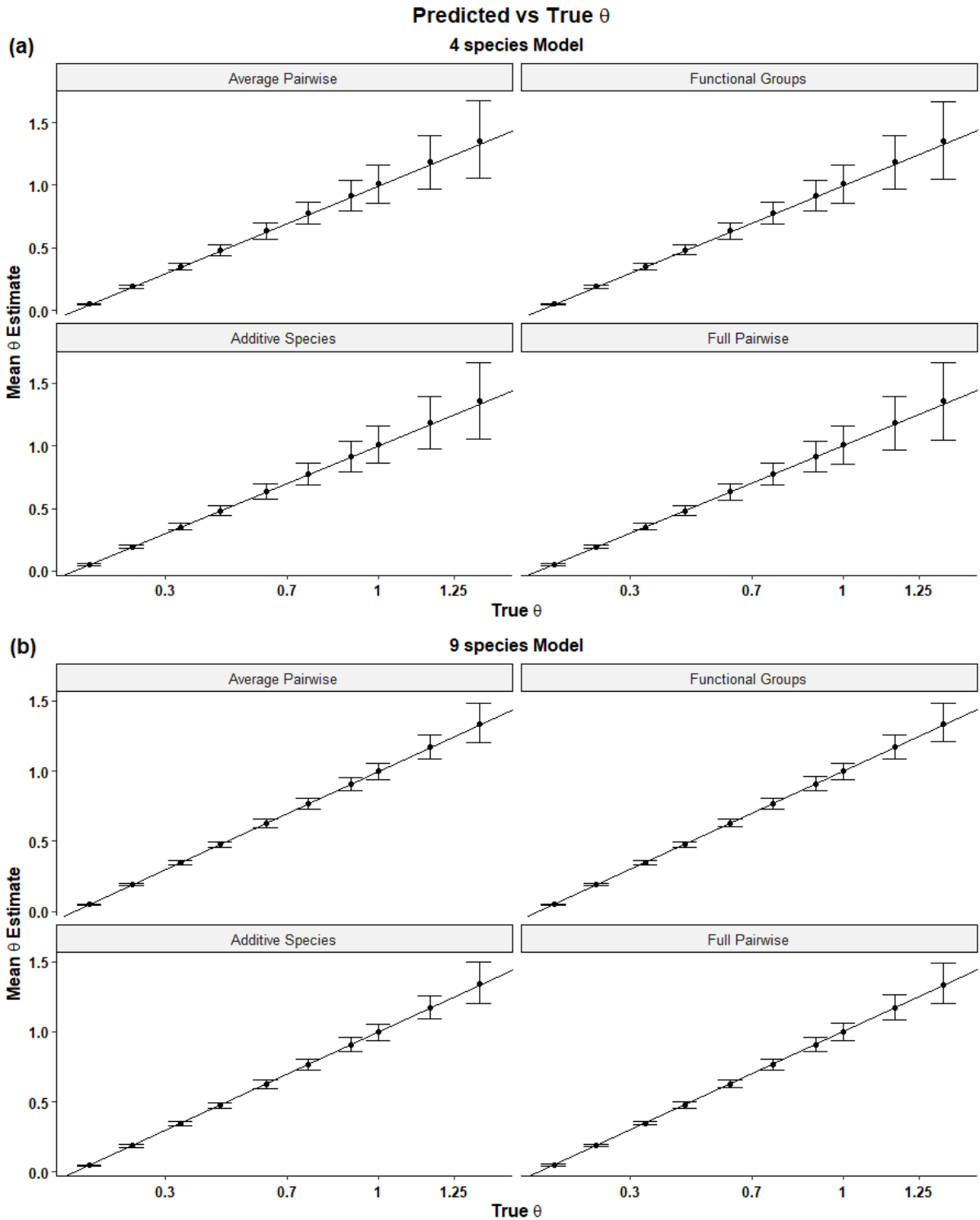

Figure 4. Mean estimated $\theta$ vs true $\theta$: (a) For the four-species model and (b) For the nine-species model. The black line in the centre is the x=y line. The $\theta$ estimates of each of the four models (average pairwise, functional group, additive species, and full pairwise) for the different $\theta$ values across the 1000 realizations (200 simulations x 5 $\sigma$ values) are averaged and represented as points. The corresponding bands around each point give the 95% dispersion of the respective estimate of $\theta$ (calculated using the 2.5% and 97.5% percentiles of the estimates of $\theta$). In each case, the true underling model was the full pairwise interactions model.

Table 4. Simulation study results: The mean, standard deviation, coverage (conditional on convergence), and distribution of $\theta$ estimates across the 200 realizations for $\sigma = 1$ of each unique $\theta$ for each of the four different DI models tested (average pairwise, functional group, additive species, and full pairwise). Similar results were observed for other $\sigma$ values (0.8, 0.9, 1.1 and 1.2), but are not shown.

(a) Four-species model: These estimates were generated with the true underlying model being the full pairwise model with identity effects 5, 6, 7, and 3 for species $SP_1$, $SP_2$, $SP_3$, and $SP_4$ respectively and interactions being simulated from normal distribution N(8,16). The functional grouping structure assumed was $SP_1$ and $SP_2$ being in FG1 and $SP_3$ and $SP_4$ being in FG2.

| | Model | | | | | | | | | | | | | | | | |
|---|---|---|---|---|---|---|---|---|---|---|---|---|---|---|---|---|---|
| | Average Pairwise | | | | Functional Group | | | | Additive Species | | | | Full Pairwise | | | | |
| True $\theta$ | Mean Est | SD Est | Coverage | Distribution | Mean Est | SD Est | Coverage | Distribution | Mean Est | SD Est | Coverage | Distribution | Mean Est | SD Est | Coverage | Distribution |
| 0.05 | 0.050 | 0.0032 | 0.975 | | 0.050 | 0.0032 | 0.91 | | 0.050 | 0.0032 | 0.86 | | 0.050 | 0.0032 | 0.37 | |
| 0.19 | 0.191 | 0.0064 | 1 | | 0.191 | 0.0064 | 0.995 | | 0.191 | 0.0064 | 0.995 | | 0.190 | 0.0064 | 0.915 | |
| 0.35 | 0.352 | 0.0129 | 1 | | 0.351 | 0.0129 | 0.99 | | 0.351 | 0.0128 | 0.99 | | 0.351 | 0.0128 | 0.895 | |
| 0.48 | 0.483 | 0.0205 | 0.995 | | 0.483 | 0.0205 | 0.99 | | 0.483 | 0.0204 | 0.985 | | 0.482 | 0.0203 | 0.91 | |
| 0.63 | 0.634 | 0.0319 | 0.985 | | 0.634 | 0.0317 | 0.975 | | 0.634 | 0.0316 | 0.975 | | 0.634 | 0.0315 | 0.925 | |
| 0.77 | 0.776 | 0.0450 | 0.975 | | 0.776 | 0.0448 | 0.97 | | 0.776 | 0.0445 | 0.97 | | 0.775 | 0.0445 | 0.945 | |
| 0.91 | 0.918 | 0.0614 | 0.965 | | 0.918 | 0.0610 | 0.96 | | 0.918 | 0.0606 | 0.96 | | 0.917 | 0.0605 | 0.945 | |
| 1 | 1.009 | 0.0742 | 0.965 | | 1.009 | 0.0737 | 0.965 | | 1.010 | 0.0733 | 0.96 | | 1.009 | 0.0732 | 0.935 | |
| 1.17 | 1.183 | 0.1065 | 0.955 | | 1.184 | 0.1060 | 0.95 | | 1.185 | 0.1055 | 0.94 | | 1.184 | 0.1054 | 0.93 | |
| 1.33 | 1.351 | 0.1535 | 0.945 | | 1.352 | 0.1534 | 0.94 | | 1.355 | 0.1535 | 0.945 | | 1.354 | 0.1535 | 0.94 | |

(b) Nine-species model: These estimates were generated with the true underlying model being the full pairwise model with identity effects 3, 4, 4, 7, 4, 6, 8, 9, and 2 for species $SP_1$, $SP_2$, $SP_3$, ..., $SP_9$ respectively and interactions being simulated from normal distribution N(10,16). The functional grouping structure assumed was $SP_1$, $SP_2$, $SP_3$, $SP_4$, and $SP_5$ being in FG1, $SP_6$ and $SP_7$ being in FG2, and $SP_8$ and $SP_9$ being in FG3.

| | Model | | | | | | | | | | | | | | | |
|---|---|---|---|---|---|---|---|---|---|---|---|---|---|---|---|---|
| | Average Pairwise | | | | Functional Group | | | | Additive Species | | | | Full Pairwise | | | |
| True $\theta$ | Mean Est | SD Est | Coverage | Distribution | Mean Est | SD Est | Coverage | Distribution | Mean Est | SD Est | Coverage | Distribution | Mean Est | SD Est | Coverage | Distribution |
| 0.05 | 0.05 | 0.002 | 1 | | 0.050 | 0.0026 | 1 | | 0.048 | 0.0026 | 1 | | 0.05 | 0.0026 | 0.2 | |
| 0.19 | 0.19 | 0.0042 | 1 | | 0.19 | 0.0042 | 1 | | 0.187 | 0.0042 | 1 | | 0.19 | 0.0042 | 0.162 | |
| 0.35 | 0.349 | 0.0069 | 1 | | 0.35 | 0.0069 | 1 | | 0.346 | 0.0070 | 1 | | 0.349 | 0.0069 | 0.65 | |
| 0.48 | 0.479 | 0.0100 | 1 | | 0.48 | 0.0099 | 1 | | 0.476 | 0.0101 | 1 | | 0.479 | 0.0099 | 0.88 | |
| 0.63 | 0.629 | 0.0144 | 1 | | 0.629 | 0.0143 | 1 | | 0.626 | 0.0147 | 1 | | 0.629 | 0.0144 | 0.905 | |
| 0.77 | 0.769 | 0.0193 | 1 | | 0.77 | 0.0193 | 1 | | 0.767 | 0.0196 | 0.995 | | 0.77 | 0.0194 | 0.92 | |
| 0.91 | 0.91 | 0.0252 | 0.99 | | 0.910 | 0.0252 | 0.99 | | 0.909 | 0.0254 | 0.99 | | 0.910 | 0.0253 | 0.915 | |
| 1 | 1.001 | 0.0300 | 0.99 | | 1.001 | 0.0299 | 0.99 | | 1.000 | 0.0301 | 0.99 | | 1.001 | 0.0300 | 0.915 | |
| 1.17 | 1.173 | 0.0435 | 0.975 | | 1.173 | 0.0433 | 0.975 | | 1.173 | 0.0436 | 0.97 | | 1.173 | 0.0437 | 0.91 | |
| 1.33 | 1.339 | 0.0682 | 0.945 | | 1.339 | 0.0679 | 0.95 | | 1.339 | 0.0700 | 0.935 | | 1.337 | 0.0695 | 0.895 | |

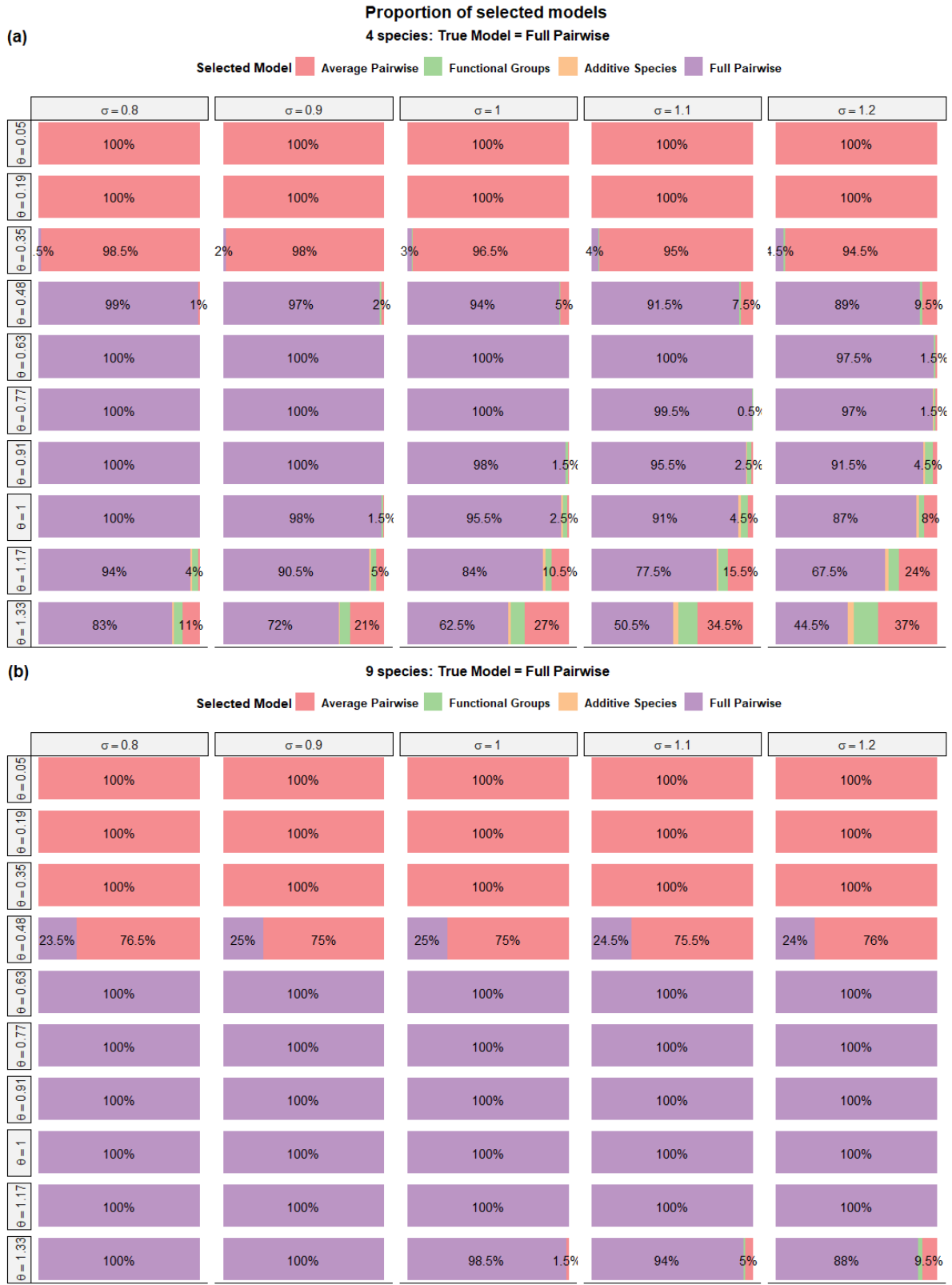

Figure 5. Efficacy of model selection procedure 'a': (a) For the four-species model and (b) for the nine-species model. The proportion of times the different GDI models are selected by the model selection procedure where we first select the appropriate interaction structure (assuming $\theta = 1$) using AIC as selection criteria and then estimate $\theta$ and test its inclusion for the selected interaction structure for the 200 simulations across each unique $\theta - \sigma$ combination. The true underlying interaction structure here is the full pairwise model.

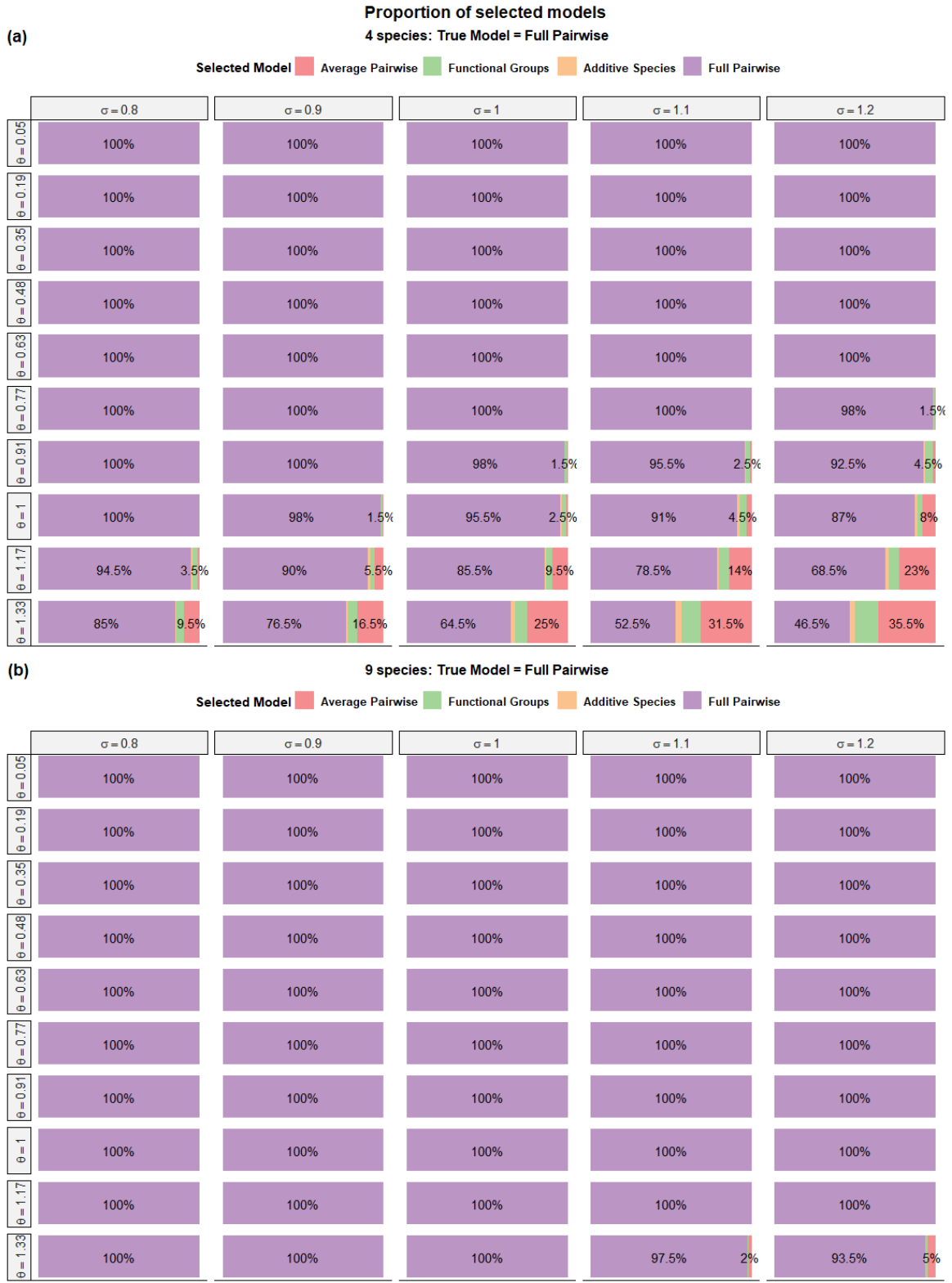

Figure 6. Efficacy of model selection procedure 'b': (a) For the four-species model and (b) for the nine-species model. The proportion of times the different GDI models are selected across the 200 simulations for each unique $\theta - \sigma$ combination when we first estimate $\theta$ for the simplest interaction structure and use the $\theta$ estimate from that model to fit and estimate the other models with different interaction structures and then select the best model using AIC as the selection metric. The true underlying interaction structure here is the full pairwise model.

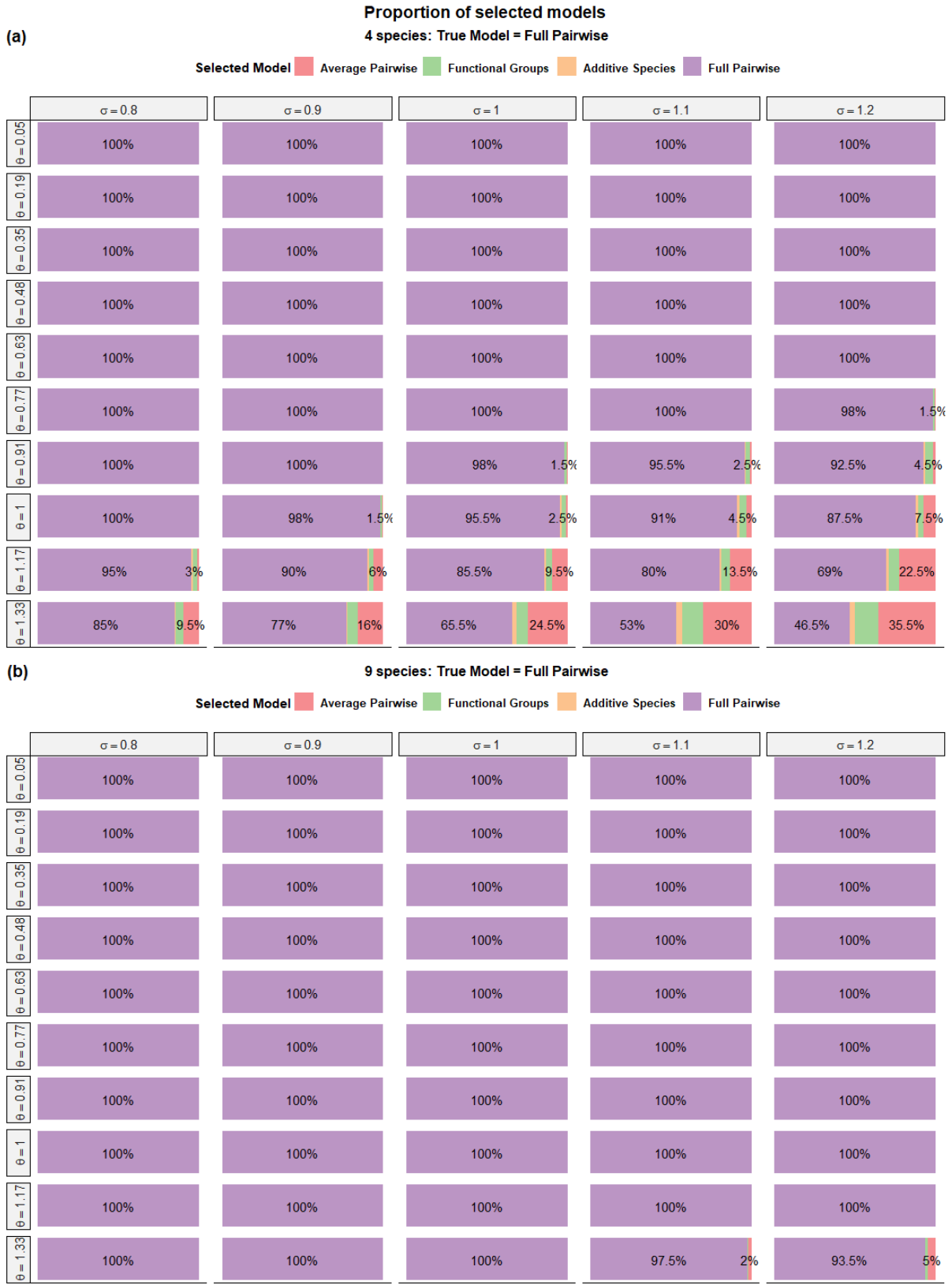

Figure 7. Efficacy of model selection procedure 'c': (a) For the four-species model and (b) for the nine-species model. The proportion of times the different GDI models are selected across the 200 simulations for each unique $\theta - \sigma$ combination when we estimate and test $\theta$ for each interaction structure and then select the best model using AIC as the selection metric. The true underlying interaction structure here is the full pairwise model.

**Discussion:**

Our simulation study showed that the estimator of $\theta$ is unbiased, with the estimate of $\theta$ being robust across the different interaction structures in GDI models. This robustness of $\theta$ is consistent across a wide range of $\theta$ values (0.05 up to 1.3) and a variety of different scenarios, including changes to the number of species (up to 72), true underlying interaction structure, functional grouping of species, presence of experimental structures, and crossing of interaction terms with experimental structures.

The results of our study give conclusive evidence for the robustness of $\theta$ and thus help us in deducing a model fitting procedure for GDI models which is parsimonious and informative. We recommend first estimating $\theta$ and testing its inclusion for either the most complex (full pairwise) or the simplest interaction structure (average pairwise) and then using that estimate of $\theta$ to fit the remaining interaction structures and perform model selection. This approach is desirable due to the speed with which we can converge to an appropriate GDI model and is used by the `autoDI` function in version 1.2. of the `DImodels` R package to perform model selection (note that the `autoDI` function in earlier versions 1.0 and 1.1 of the `DImodels` R package did not use this recommended method of model selection). There is precedence for fixing a model parameter to a specific value when performing model selection in wider statistics. For example, in negative binomial models, the dispersion parameter may be fixed when testing for model effects; in generalized linear mixed models including observational-level random effects (e.g. Poisson-normal, see [Demétrio et al., 2014](#)), the variance of the random effects may also be fixed when testing fixed effects.

This robustness in $\theta$ estimation does rely on a few prerequisites being satisfied. Firstly, there should be adequate coverage of the species around the simplex space. If the communities are not well-spread across the simplex space, then there may not be enough information in the data and hence the estimation of θ could fail for certain models. Further, if there is lack-of-fit in the models or if all, or a majority, of the interaction terms aren't significant then this would have trickle-down effects on $\theta$ and its estimation would be affected, resulting in estimates which are considerably different across the different GDI models while at the same time being far off from the true underlying value of $\theta$. Thus, it is important to check for any data issues and model fit before finalising model selection. Examples with such issues are highlighted in Appendix C. In this study, we have examined over 300,000 datasets, which gives assurance in the reliability of our results. Our study assumed a single $\theta$ parameter across all interaction terms; there is scope for allowing different $\theta$'s for each interaction term in GDI models, perhaps in a multivariate or repeated measures setting with complex variance structures.

**Conclusion:**

Our simulation study has shown that for our experimental designs, $\theta$ estimation is invariant to different interaction structures and that the most efficient model selection procedure for GDI models is to first estimate $\theta$ and test for its significance, and then use that estimate to fit the different interaction structures and finally select the most appropriate structure using selection criteria.

**Acknowledgements:**

All authors were supported by the Science Foundation Ireland Frontiers for the Future programme, grant number 19/FFP/6888 award to CB.

**Authors' Contributions:**

CB, JC, and RV conceived the ideas. RV led the coding with contributions from LB and RAM. RV and CB wrote the initial draft of the paper. LB, JC, and RAM each contributed to the writing of the subsequent drafts of the paper.

# Appendix A

| Richness | Species | | | |
|---|---|---|---|---|
| | p1 | p2 | p3 | p4 |
| 1 | 1 | | | |
| 1 | | 1 | | |
| 1 | | | 1 | |
| 1 | | | | 1 |
| 2 | 0.5 | 0.5 | | |
| 2 | 0.5 | | 0.5 | |
| 2 | 0.5 | | | 0.5 |
| 2 | | 0.5 | 0.5 | |
| 2 | | 0.5 | | 0.5 |
| 2 | | | 0.5 | 0.5 |
| 3 | 0.333333 | 0.333333 | 0.333333 | |
| 3 | 0.333333 | 0.333333 | | 0.333333 |
| 3 | 0.333333 | | 0.333333 | 0.333333 |
| 3 | | 0.333333 | 0.333333 | 0.333333 |
| 4 | 0.7 | 0.1 | 0.1 | 0.1 |
| 4 | 0.4 | 0.4 | 0.1 | 0.1 |
| 4 | 0.4 | 0.2 | 0.2 | 0.2 |
| 4 | 0.4 | 0.1 | 0.4 | 0.1 |
| 4 | 0.4 | 0.1 | 0.1 | 0.4 |
| 4 | 0.3 | 0.3 | 0.3 | 0.1 |
| 4 | 0.3 | 0.3 | 0.1 | 0.3 |
| 4 | 0.3 | 0.1 | 0.3 | 0.3 |
| 4 | 0.25 | 0.25 | 0.25 | 0.25 |
| 4 | 0.2 | 0.4 | 0.2 | 0.2 |
| 4 | 0.2 | 0.2 | 0.4 | 0.2 |
| 4 | 0.2 | 0.2 | 0.2 | 0.4 |
| 4 | 0.1 | 0.7 | 0.1 | 0.1 |
| 4 | 0.1 | 0.4 | 0.4 | 0.1 |
| 4 | 0.1 | 0.4 | 0.1 | 0.4 |
| 4 | 0.1 | 0.3 | 0.3 | 0.3 |
| 4 | 0.1 | 0.1 | 0.7 | 0.1 |
| 4 | 0.1 | 0.1 | 0.4 | 0.4 |
| 4 | 0.1 | 0.1 | 0.1 | 0.7 |
| 4 | 0.9 | 0.033333 | 0.033333 | 0.033333 |
| 4 | 0.033333 | 0.9 | 0.033333 | 0.033333 |
| 4 | 0.033333 | 0.033333 | 0.9 | 0.033333 |
| 4 | 0.033333 | 0.033333 | 0.033333 | 0.9 |

Figure A1. Design for four species model: The experimental design for the four species model is shown. Each of these communities was replicated three times.

| Richness | Species | | | | | | | | |
|---|---|---|---|---|---|---|---|---|---|
|  | p1 | p2 | p3 | p4 | p5 | p6 | p7 | p8 | p9 |
| 1 |  |  |  |  |  |  |  |  | 1 |
| 1 |  |  |  |  |  |  |  | 1 |  |
| 1 |  |  |  |  |  |  | 1 |  |  |
| 1 |  |  |  |  |  | 1 |  |  |  |
| 1 |  |  |  |  | 1 |  |  |  |  |
| 1 |  |  |  | 1 |  |  |  |  |  |
| 1 |  |  | 1 |  |  |  |  |  |  |
| 1 |  | 1 |  |  |  |  |  |  |  |
| 1 | 1 |  |  |  |  |  |  |  |  |
| 2 |  |  |  |  |  |  |  | 0.5 | 0.5 |
| 2 |  |  |  |  |  |  | 0.5 |  | 0.5 |
| 2 |  |  |  |  |  | 0.5 |  |  | 0.5 |
| 2 |  |  |  |  | 0.5 |  |  |  | 0.5 |
| 2 |  |  |  | 0.5 |  |  |  |  | 0.5 |
| 2 |  |  | 0.5 |  |  |  |  |  | 0.5 |
| 2 |  | 0.5 |  |  |  |  |  |  | 0.5 |
| 2 | 0.5 |  |  |  |  |  |  |  | 0.5 |
| 2 |  |  |  |  |  |  | 0.5 | 0.5 |  |
| 2 |  |  |  |  |  | 0.5 |  | 0.5 |  |
| 2 |  |  |  |  | 0.5 |  |  | 0.5 |  |
| 2 |  |  |  | 0.5 |  |  |  | 0.5 |  |
| 2 |  |  | 0.5 |  |  |  |  | 0.5 |  |
| 2 |  | 0.5 |  |  |  |  |  | 0.5 |  |
| 2 | 0.5 |  |  |  |  |  |  | 0.5 |  |
| 2 |  |  |  |  |  | 0.5 | 0.5 |  |  |
| 2 |  |  |  |  | 0.5 |  | 0.5 |  |  |
| 2 |  |  |  | 0.5 |  |  | 0.5 |  |  |
| 2 |  |  | 0.5 |  |  |  | 0.5 |  |  |
| 2 |  | 0.5 |  |  |  |  | 0.5 |  |  |
| 2 | 0.5 |  |  |  |  |  | 0.5 |  |  |
| 2 |  |  |  |  | 0.5 | 0.5 |  |  |  |
| 2 |  |  |  | 0.5 |  | 0.5 |  |  |  |
| 2 |  |  | 0.5 |  |  | 0.5 |  |  |  |
| 2 |  | 0.5 |  |  |  | 0.5 |  |  |  |
| 2 | 0.5 |  |  |  |  | 0.5 |  |  |  |
| 2 |  |  |  | 0.5 | 0.5 |  |  |  |  |
| 2 |  |  | 0.5 |  | 0.5 |  |  |  |  |
| 2 |  | 0.5 |  |  | 0.5 |  |  |  |  |
| 2 | 0.5 |  |  |  | 0.5 |  |  |  |  |
| 2 |  |  | 0.5 | 0.5 |  |  |  |  |  |
| 2 |  | 0.5 |  | 0.5 |  |  |  |  |  |
| 2 | 0.5 |  |  | 0.5 |  |  |  |  |  |
| 2 |  | 0.5 | 0.5 |  |  |  |  |  |  |
| 2 | 0.5 |  | 0.5 |  |  |  |  |  |  |
| 2 | 0.5 | 0.5 |  |  |  |  |  |  |  |

| # | 1 | 2 | 3 | 4 | 5 | 6 | 7 | 8 | 9 |
|---|---|---|---|---|---|---|---|---|---|
| 3 |  |  |  | 0.333333 |  |  |  | 0.333333 | 0.333333 |
| 3 |  | 0.333333 |  |  |  |  |  | 0.333333 | 0.333333 |
| 3 |  |  |  |  |  | 0.333333 | 0.333333 |  | 0.333333 |
| 3 |  |  |  | 0.333333 |  |  | 0.333333 |  | 0.333333 |
| 3 | 0.333333 |  |  |  |  | 0.333333 |  |  | 0.333333 |
| 3 |  |  | 0.333333 |  | 0.333333 |  |  |  | 0.333333 |
| 3 |  | 0.333333 |  |  | 0.333333 |  |  |  | 0.333333 |
| 3 | 0.333333 |  | 0.333333 |  |  |  |  |  | 0.333333 |
| 3 |  |  |  |  |  | 0.333333 | 0.333333 | 0.333333 |  |
| 3 |  | 0.333333 |  |  |  |  | 0.333333 | 0.333333 |  |
| 3 |  | 0.333333 |  |  |  | 0.333333 |  | 0.333333 |  |
| 3 |  |  |  | 0.333333 | 0.333333 |  |  | 0.333333 |  |
| 3 | 0.333333 |  |  |  | 0.333333 |  |  | 0.333333 |  |
| 3 | 0.333333 |  | 0.333333 |  |  |  |  | 0.333333 |  |
| 3 |  |  |  | 0.333333 | 0.333333 |  | 0.333333 |  |  |
| 3 | 0.333333 |  |  |  | 0.333333 |  | 0.333333 |  |  |
| 3 |  | 0.333333 | 0.333333 |  |  |  | 0.333333 |  |  |
| 3 | 0.333333 | 0.333333 |  |  |  |  | 0.333333 |  |  |
| 3 |  |  | 0.333333 |  | 0.333333 | 0.333333 |  |  |  |
| 3 |  | 0.333333 |  |  | 0.333333 | 0.333333 |  |  |  |
| 3 |  |  | 0.333333 | 0.333333 |  | 0.333333 |  |  |  |
| 3 | 0.333333 |  |  | 0.333333 |  | 0.333333 |  |  |  |
| 3 |  | 0.333333 | 0.333333 | 0.333333 |  |  |  |  |  |
| 3 | 0.333333 | 0.333333 |  | 0.333333 |  |  |  |  |  |
| 4 |  | 0.25 |  |  |  |  | 0.25 | 0.25 | 0.25 |
| 4 | 0.25 |  |  |  | 0.25 |  |  | 0.25 | 0.25 |
| 4 |  | 0.25 | 0.25 |  |  |  |  | 0.25 | 0.25 |
| 4 |  |  |  |  | 0.25 | 0.25 | 0.25 |  | 0.25 |
| 4 | 0.25 |  | 0.25 |  |  |  | 0.25 |  | 0.25 |
| 4 |  |  | 0.25 | 0.25 |  | 0.25 |  |  | 0.25 |
| 4 |  | 0.25 |  | 0.25 |  | 0.25 |  |  | 0.25 |
| 4 | 0.25 |  |  | 0.25 | 0.25 |  |  |  | 0.25 |
| 4 |  |  |  |  | 0.25 | 0.25 | 0.25 | 0.25 |  |
| 4 |  |  | 0.25 | 0.25 |  |  | 0.25 | 0.25 |  |
| 4 | 0.25 |  |  | 0.25 |  | 0.25 |  | 0.25 |  |
| 4 | 0.25 | 0.25 |  |  |  | 0.25 |  | 0.25 |  |
| 4 |  |  | 0.25 | 0.25 | 0.25 |  |  | 0.25 |  |
| 4 | 0.25 |  | 0.25 |  |  | 0.25 | 0.25 |  |  |
| 4 |  | 0.25 |  | 0.25 | 0.25 |  | 0.25 |  |  |
| 4 | 0.25 | 0.25 |  | 0.25 |  |  | 0.25 |  |  |
| 4 |  | 0.25 | 0.25 |  | 0.25 | 0.25 |  |  |  |
| 4 | 0.25 | 0.25 | 0.25 |  | 0.25 |  |  |  |  |
| 6 |  |  | 0.166667 |  | 0.166667 | 0.166667 | 0.166667 | 0.166667 | 0.166667 |
| 6 | 0.166667 | 0.166667 |  |  | 0.166667 |  | 0.166667 | 0.166667 | 0.166667 |
| 6 | 0.166667 |  | 0.166667 | 0.166667 |  |  | 0.166667 | 0.166667 | 0.166667 |
| 6 | 0.166667 |  |  | 0.166667 | 0.166667 | 0.166667 |  | 0.166667 | 0.166667 |
| 6 |  | 0.166667 | 0.166667 | 0.166667 |  | 0.166667 |  | 0.166667 | 0.166667 |
| 6 |  | 0.166667 |  | 0.166667 | 0.166667 | 0.166667 | 0.166667 |  | 0.166667 |
| 6 | 0.166667 | 0.166667 | 0.166667 |  |  | 0.166667 | 0.166667 |  | 0.166667 |
| 6 | 0.166667 | 0.166667 | 0.166667 | 0.166667 | 0.166667 |  |  |  | 0.166667 |
| 6 | 0.166667 | 0.166667 |  | 0.166667 |  | 0.166667 | 0.166667 | 0.166667 |  |
| 6 |  | 0.166667 | 0.166667 | 0.166667 | 0.166667 |  | 0.166667 | 0.166667 |  |
| 6 | 0.166667 | 0.166667 | 0.166667 |  | 0.166667 | 0.166667 |  | 0.166667 |  |
| 6 | 0.166667 |  | 0.166667 | 0.166667 | 0.166667 | 0.166667 | 0.166667 |  |  |
| 9 | 0.111111 | 0.111111 | 0.111111 | 0.111111 | 0.111111 | 0.111111 | 0.111111 | 0.111111 | 0.111111 |

Figure A2. Design for nine species model: The experimental design for the nine species model is shown.

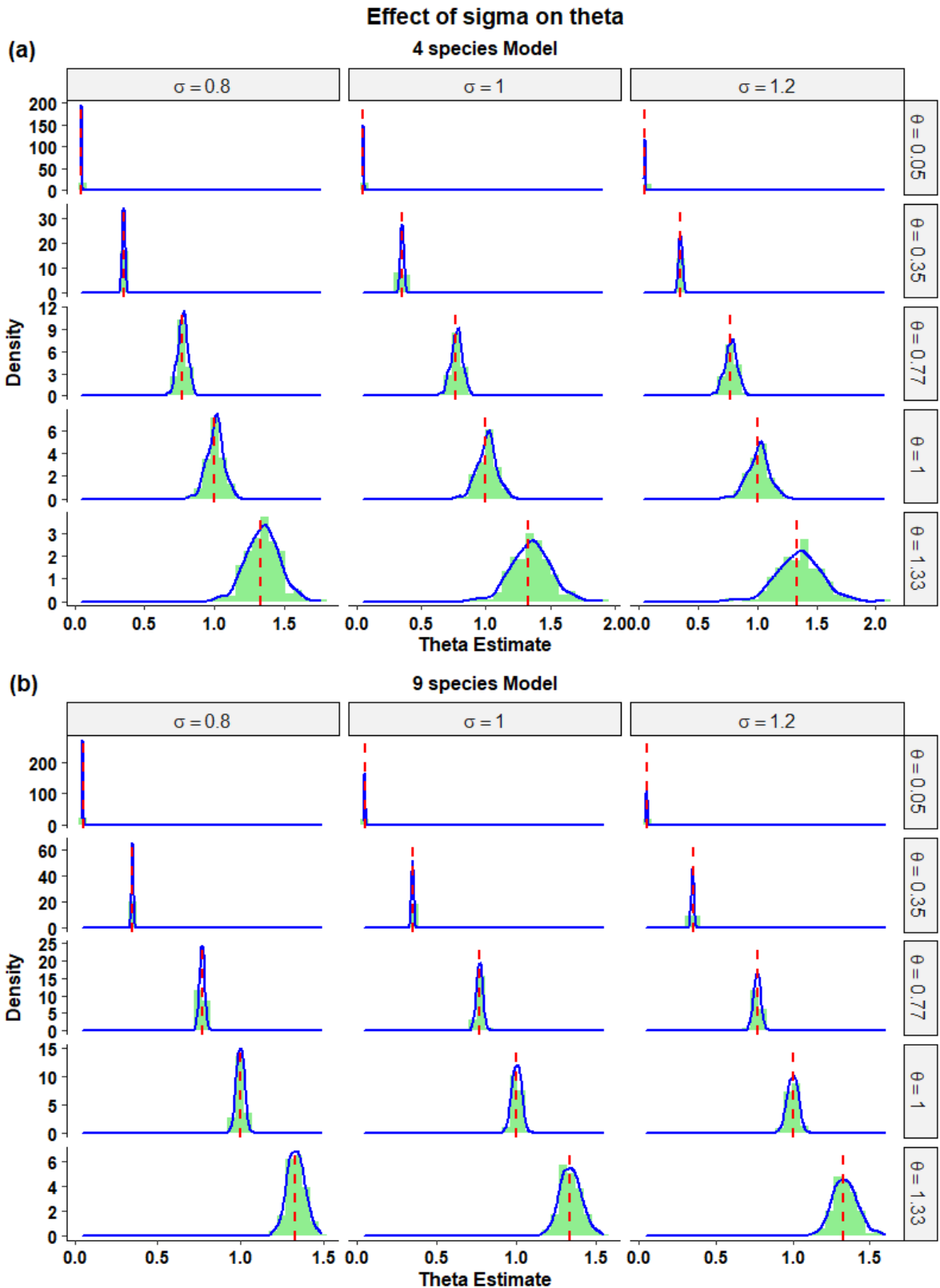

Figure A3. Effect of changing σ on the robustness of θ estimation: The distribution of θ estimates for the average pairwise model across the 200 simulations for a select range of θ and σ values. The blue curve is the density curve while the dashed red line is the true value of θ. The overall results did not change for the other interaction structure models or for the other values of θ and σ and are not shown.

**Unusual coverages for the confidence intervals**

The coverages for the 95% confidence intervals (CI) were not always near the 0.95 mark particularly for low values of $\theta$ near 0. This was partially because we were using a singular $\theta$ value across all the different interaction structures. This resulted in different CIs for the interaction structures even though the $\theta$ estimate was the same. The CI was inversely proportional to the number of number of terms in the model involving $\theta$ and hence the average pairwise model had a widest CI whilst the full pairwise model had the narrowest CI, with the other model have CI ranging between the two. This could explain why the full pairwise model had such low coverage values for low values of $\theta$ whilst the other interaction structures had coverage values near 1. Table A1 shows an example of this. This effect occurred across the entire range of $\theta$ values but was particularly evident near low values of $\theta$ as the intervals here were too precise and any minor changes in the intervals resulted in drastic changes in the coverages. A possible solution for this could be to test for different values of $\theta$ for each interaction term in the model.

Table A1. All four interaction structures are fit to a specific dataset from the nine species example (true $\theta$ = 0.05, $\sigma$ = 1, seed == 1005) and the $\theta$ estimate, lower and upper CI and the width of the CI is presented.

| Model | $\theta$ Estimate | Lower CI | Upper CI | CI Width |
|---|---|---|---|---|
| Average Pairwise | 0.0522 | 0.0177 | 0.0874 | 0.0697 |
| Functional Group | 0.052 | 0.0201 | 0.0855 | 0.0654 |
| Additive Species | 0.0502 | 0.0228 | 0.0772 | 0.0544 |
| Full Pairwise | 0.0522 | 0.0575 | 0.061 | 0.0035 |

Another reason for the abnormal coverages was due to convergence problems near the boundary. Figure A4 shows a comparison between the likelihood curves for an example with no convergence problems and an example with convergence problems where it wasn't possible to estimate a CI. Under normal situations the CI is defined by the two points at which the likelihood curve intersects the X-axis (Figure A4 (a)) but because we restrict $\theta$ to have only positive values, it can sometimes happen that the likelihood curve intersects the x-axis at only one point in our region of interest (with the other point being on the negative side of the y-axis) as shown in Figure A4 (b). This results in convergence problems and thus the CI can't be estimated.

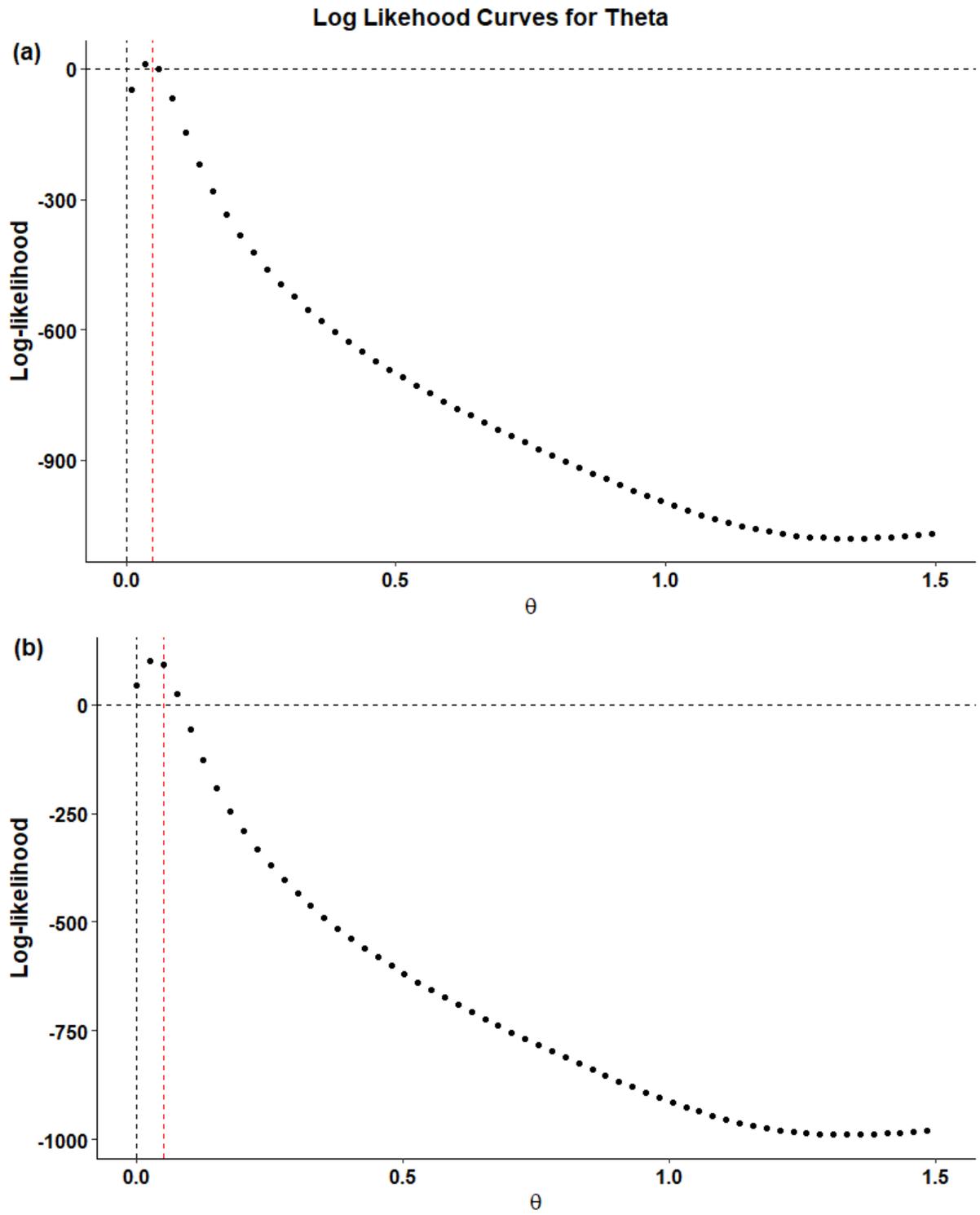

Figure A4. Likelihood curve for $\theta$ in two different scenarios: (a) Normal scenario with convergence and (b) Condition where convergence fails. The dotted curve represents the likelihood curve, while the dashed red line is the true value of $\theta$.

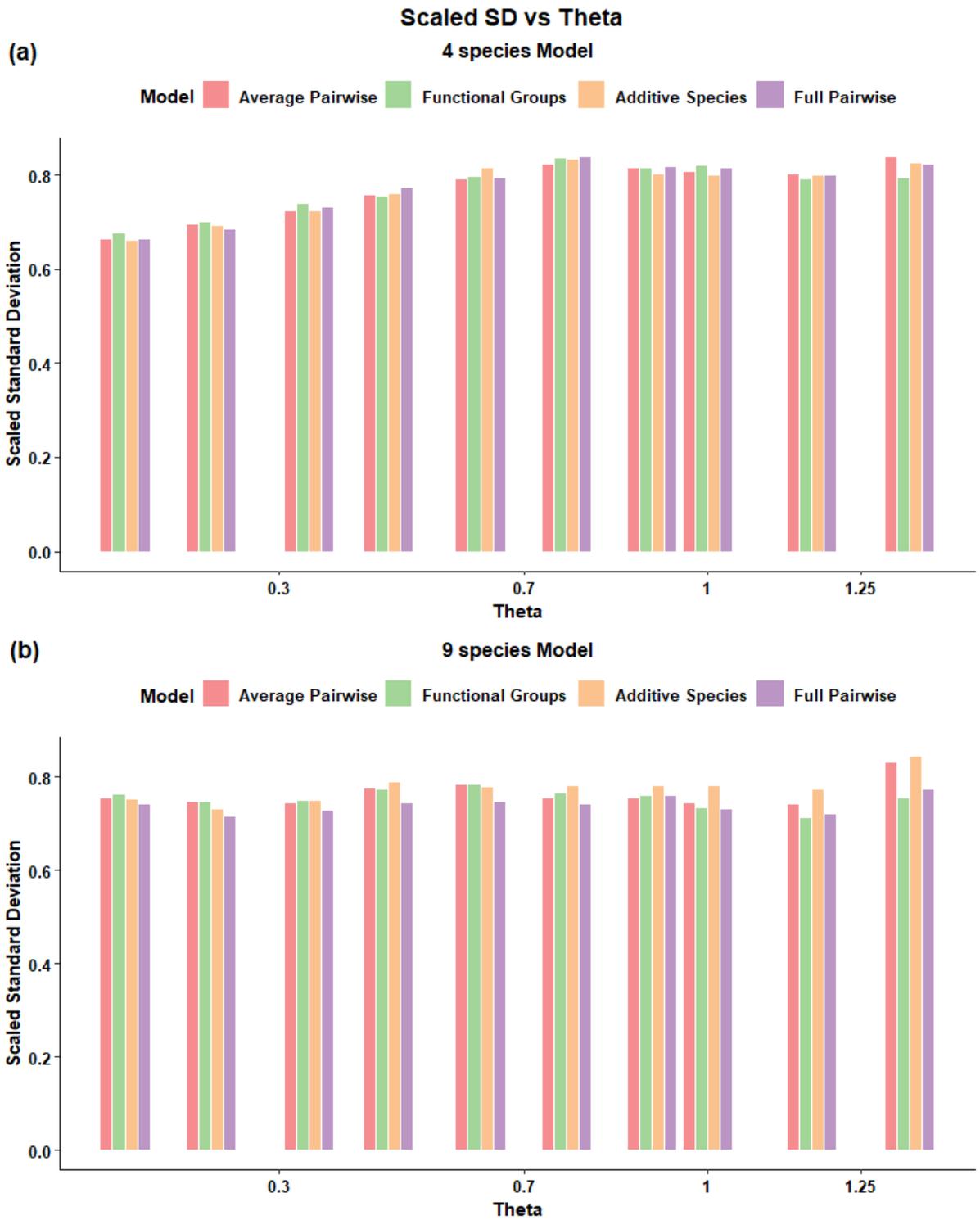

Figure A5. Scaled standard deviation vs $\theta$: (a) For the 4 species model and (b) for the nine species model. The bars represent the scaled standard deviations which are obtained by dividing the standard deviations by the inter-quartile range of each unique $\theta$-model combination. The bars are coloured for the 4 different models (average pairwise, functional group, additive species, and full pairwise).

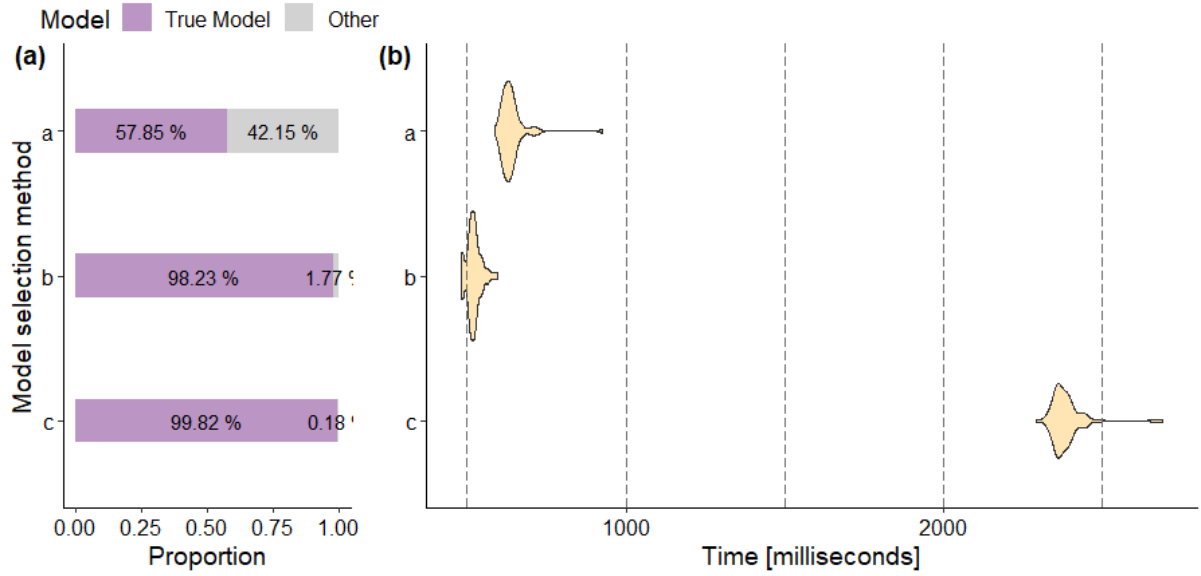

Figure A6. Efficacy and computation time comparison between the three model selection procedures for GDI models tested: (a) The proportion of times the true model is selected across the 10,000 (10 $\theta$ x 5 $\sigma$ x 200 simulations) datasets for the nine species example using AIC as the selection metric. (b) The distribution of the computation time for selecting the best model for a given dataset across the three model selection procedures.

# Appendix B

# $\theta$ Reparameterization Simulation

**Background:**

Generalized Diversity-Interactions models (GDI; Connolly et al. 2013) express the biodiversity and ecosystem functioning (BEF) relationship as a function of the species identities and their interactions exponentiated with a coefficient $\theta$ to account for non-linearity. GDI model are adept at modelling the BEF relationship but suffer from a complication where the interaction effect at the centroid community is dependent on both the value of $\theta$ as well as the interaction effect coefficient $\delta$. Connolly et al. 2018 suggested a reformulation of GDI models to tackle this complication where the interaction effect is scaled by a factor of $\frac{2s^{2\theta}}{s(s-1)}$, giving the following formulation for a GDI model:

$$y = \sum_{i=1}^{S} \beta_i P_i + \frac{2s^{2\theta}}{s(s-1)} \sum_{\substack{i,j=1 \\ i<j}}^{S} \delta_{ij} (P_i P_j)^\theta + \alpha A + \epsilon \qquad Eq\ 1$$

where $s$ is the number of species in the experimental design, $P_i$ and $P_j$ are the proportions of the $i^{th}$ and $j^{th}$ species in the community respectively, $\beta_i$ is the identity effect of species $i$, the $\delta_{ij}$ parameters are the effects of the interactions between species $i$ and $j$, $A$ and $\alpha$ are vectors of experiment structures and the effects of those experimental structures respectively with $\epsilon$ being the normally distributed error term with mean 0 and variance $\sigma^2$, i.e. $\varepsilon \sim N(0, \sigma^2)$. The coefficient $\theta$ is the non-linear parameter affecting the strength of the interaction between a pair of species.

The scaling factor is chosen such that at the centroid where $P_i = 1/s$, the interaction effect $\delta$ is maximum and is independent of $\theta$. This formulation allows us to deal with evenness and the maximum diversity effect (essentially richness) as two separate phenomena where $\delta$ measures the maximum diversity effect at the equi-proportional mixture while $\theta$ measures the shape of the interactions between species. The scaling factor can be considered a measure of evenness (Kirwan et al. 2007; 2009) for situations where $\theta \neq 1$. An additional advantage of this reparameterization is that it will tend to reduce correlation between the estimates of the $\theta$ and $\delta$ coefficients.

**Methods:**

The natural question that arises next is how the estimate of $\theta$ gets affected under the new parameterization. We tested this by performing a simulation study using the same four and nine species datasets used in the paper. The identity and interaction effect coefficients were the same as mentioned in the paper and the response was calculated for ten different $\theta$ values ranging from 0.05 to 1.33 across five different $\sigma$ values (0.8 up to 1.2) for the error term with 200 datasets simulated for each unique $(\theta, \sigma)$ combination. The true underlying model was assumed to be the full pairwise model, but we fit the average pairwise interaction model for each unique dataset using both the old (Connolly et al. 2013) and new (Connolly et al. 2018) parameterizations of GDI models. The objectives of this simulation study were:

(i) To assess whether the estimate of $\theta$ differed across the two parameterizations
(ii) To analyse whether the correlation between $\theta$ and $\delta$ terms is reduced in the new parametrization

To test these objectives, the models were fit under both parameterizations and the estimates of $\delta$ and $\theta$ were obtained by linear least squares and profile likelihood respectively using the `DImodels` package (Moral et al. 2021) for each of the 200 simulated datasets across the 100 simulation settings (2 experimental designs x 10 $\theta$ values x 5 $\sigma$ values). The estimates of $\theta$ across the two parameterizations for each dataset were checked for identity and the correlation between the $\theta$ and $\delta$ term was calculated across the 200 datasets for each of the 100 simulation settings.

We also tested these objectives using non-linear least (nls) squares to estimate both $\delta$ and $\theta$ simultaneously to ensure the results were robust across different methods of estimation too. Further, various other datasets with different experimental designs, identities, and true underlying interaction structures were also tested to ensure robustness of results. Finally, other GDI models besides the average pairwise interaction model (see Table 1 in paper) were also fit to the datasets to corroborate that these findings were invariant across the different interaction structures too.

**Results:**

The results obtained from the study were similar for both the four and nine species cases. The estimate of the $\theta$ was identical across the two parameterizations. The correlation between $\theta$ and $\delta$ was found to be lower in the new parameterization. This effect was especially evident as the value of the $\theta$ increased, with the new parameterization having significantly less correlation between $\theta$ and $\delta$ compared to the old parameterization (Table B1). Splitting the results up by the five $\sigma$ values (0.8, 0.9, 1, 1.1, and 1.2) it was found that the results were invariant to a changing $\sigma$, with the only effect of $\sigma$ being an increase in the variation of the distribution of the estimates of $\theta$ and $\delta$ as the value of $\sigma$ increased. These results are summarized in Figures B1 and B2 for the four and nine species examples respectively.

Similar results were observed for estimation using non-linear least squares as well as estimation under different interactions structures.

**Conclusion:**

The estimate of $\theta$ is invariant across the two parameterizations for all the factors we tested against. The correlation between the $\delta$ terms and $\theta$ is reduced by a notable amount under the new parameterization compared to the old parameterization.

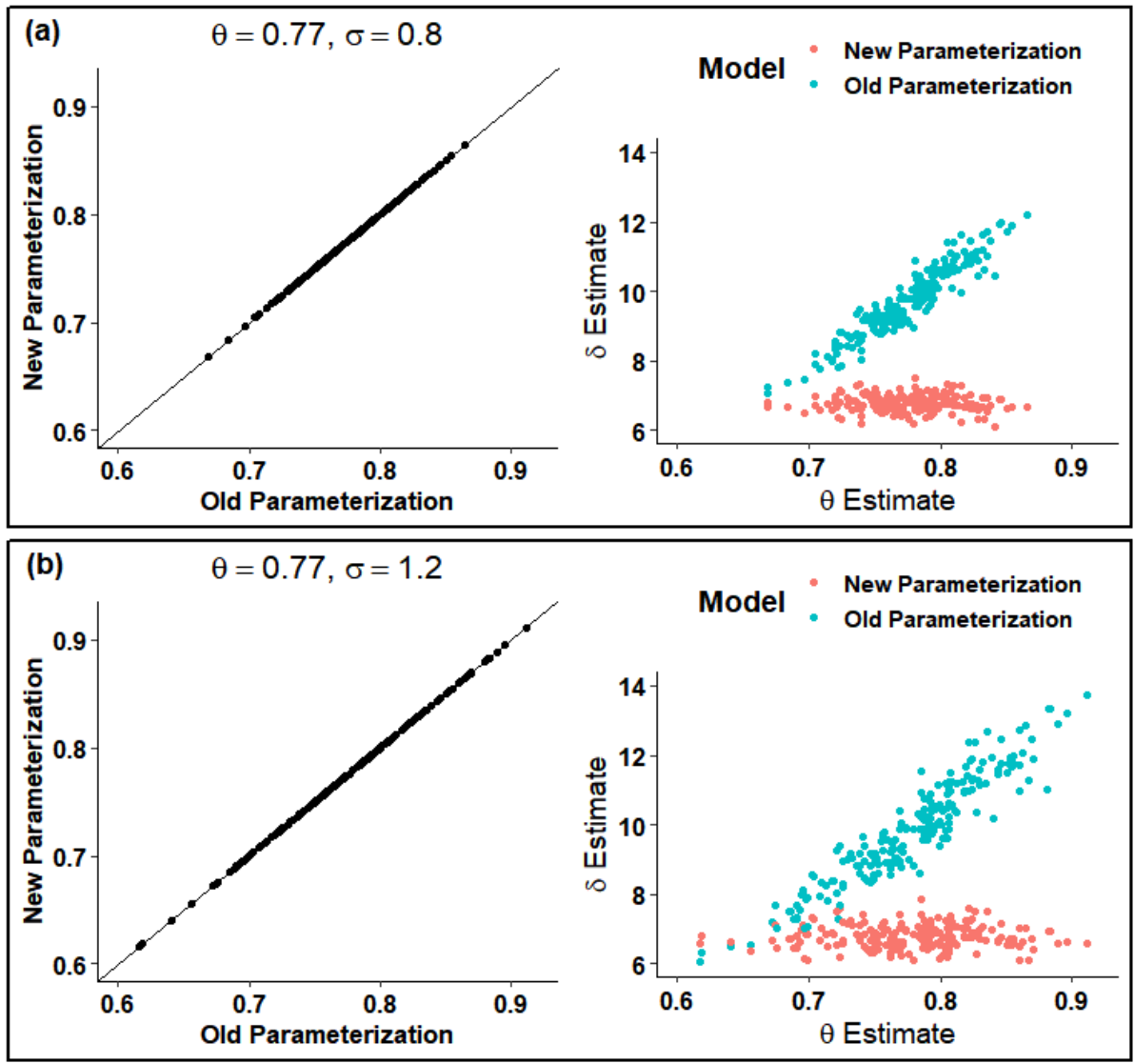

Figure B2: Identity of $\theta$ across the two parameterizations and correlation between $\theta$ and $\delta$ for the four species model: (a) Shows the estimates of $\theta$ across the new (Connolly et al. 2018) and old (Connolly et al. 2013) parameterizations on the Y and X axes respectively for the 200 simulations of the $\theta = 0.77$ and $\sigma = 0.8$ combination in the left figure with the black line being the x=y line. The figure on the right shows the scatterplot between the $\delta$ estimate and $\theta$ estimate across both the parameterizations for the same $\theta = 0.77$ and $\sigma = 0.8$ combination. (b) Shows the same results but for the $\theta = 0.77$ and $\sigma = 1.2$ combination. Similar results observed for other $(\theta, \sigma)$ combinations too.

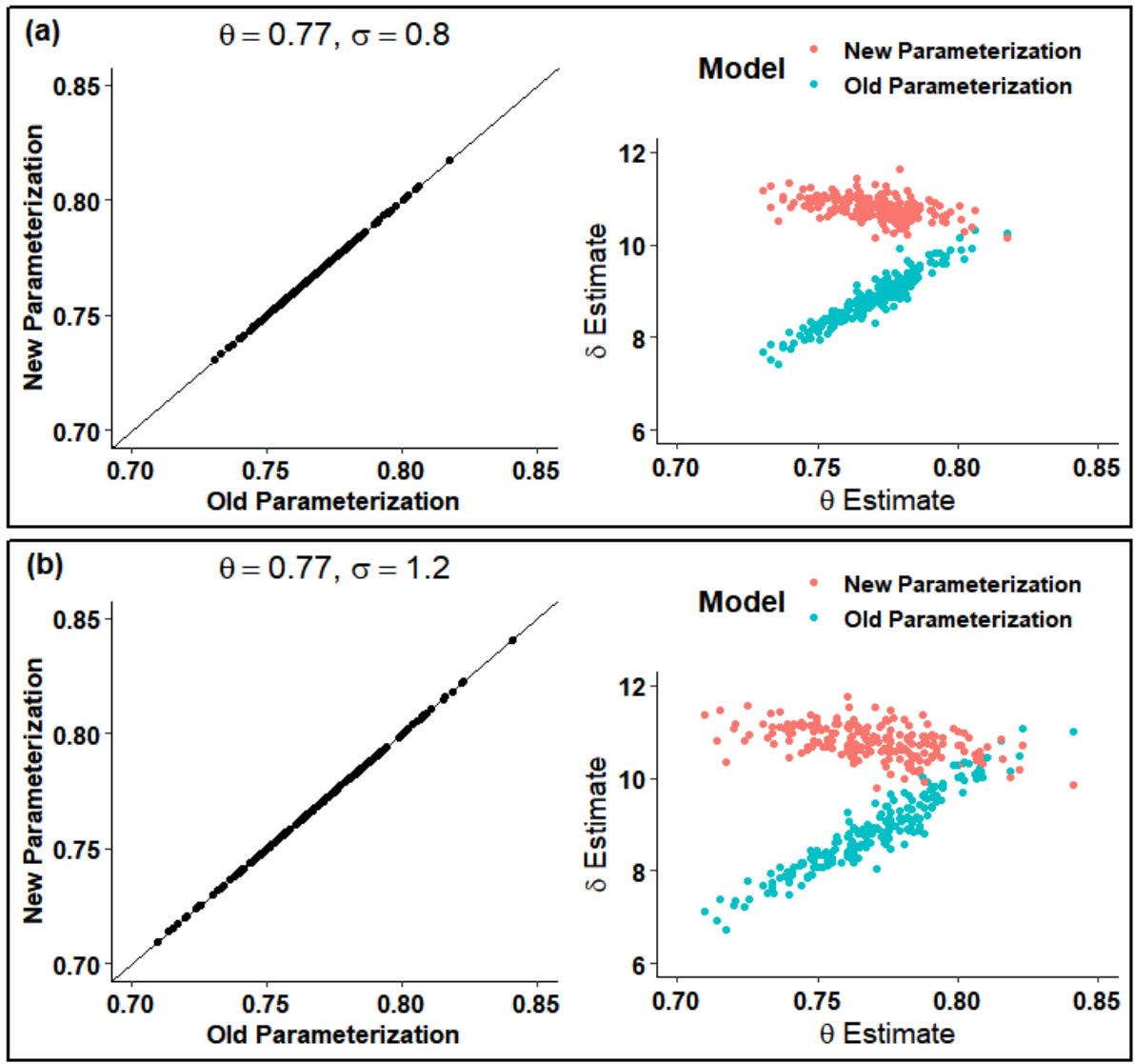

Figure B2: Identity of $\theta$ across the two parameterizations and correlation between $\theta$ and $\delta$ for the nine species model: (a) Shows the estimates of $\theta$ across the new (Connolly et al. 2018) and old (Connolly et al. 2013) parameterizations on the Y and X axes respectively for the 200 simulations of the $\theta = 0.77$ and $\sigma = 0.8$ combination in the left figure with the black line being the x=y line. The figure on the right shows the scatterplot between the $\delta$ estimate and $\theta$ estimate across both the parameterizations for the same $\theta = 0.77$ and $\sigma = 0.8$ combination. (b) Shows the same results but for the $\theta = 0.77$ and $\sigma = 1.2$ combination. Similar results observed for other $(\theta, \sigma)$ combinations too.

Table B2: Correlation between the $\theta$ and $\delta$ estimates of all 1000 realizations (200 datasets x 5 $\sigma$ values) for each $\theta$ value across the two parameterizations.

(a) For the four species example:

| $\theta$ | Model | |
|---|---|---|
| | Old Parameterization | New Parameterization |
| 0.05 | 0.9192 | 0.4539 |
| 0.19 | 0.9364 | 0.3569 |
| 0.35 | 0.9498 | 0.2715 |
| 0.48 | 0.953 | 0.199 |
| 0.63 | 0.9494 | 0.0986 |
| 0.77 | 0.9398 | -0.0123 |
| 0.91 | 0.9245 | -0.1366 |
| 1 | 0.9121 | -0.2212 |
| 1.17 | 0.8843 | -0.3795 |
| 1.33 | 0.816 | -0.4283 |

(b) For the nine species example.

| $\theta$ | Model | |
|---|---|---|
| | Old Parameterization | New Parameterization |
| 0.05 | 0.9897 | -0.6049 |
| 0.19 | 0.9875 | -0.5923 |
| 0.35 | 0.9835 | -0.5607 |
| 0.48 | 0.9778 | -0.5209 |
| 0.63 | 0.9659 | -0.4698 |
| 0.77 | 0.9442 | -0.4404 |
| 0.91 | 0.9038 | -0.4596 |
| 1 | 0.8633 | -0.5059 |
| 1.17 | 0.7563 | -0.6506 |
| 1.33 | 0.6521 | -0.7893 |

# Appendix C

## Additional simulations for testing robustness of $\theta$ and model selection efficacy of proposed model selection method

Additional simulations were performed to test the robustness of $\theta$ and model selection efficacy by varying factors like number of species, presence of experimental structures, different true underlying interaction structures as well as different structures of the function groupings. Results from three specific simulations are presented here which include a) 6 species model with true interaction structure being additive species along with a treatment effect, b) 16 species model with a specific functional grouping and tested for a different functional grouping, and c) 72 species model with average pairwise interaction structure being the true underlying interaction structure. These three simulation examples presented here cover all the different factors tested, however many more simulations were run to ensure reliability of the observed findings, the results of which haven't been presented here.

### a) 6 species model

The six species design consisted of species $SP_1$, $SP_2$, $SP_3$, $SP_4$, $SP_5$ and $SP_6$ spread across the simplex space (Cornell 2011, Kirwan et al. 2007) in a design consisting of 19 different communities comprised of six monocultures and 13 mixtures from two to six species. The design opted was the same used in Grange et al. 2021. Each community was repeated thrice and the following functional grouping structure was assumed with $SP_1$ and $SP_2$ being in functional group one, $SP_3$ and $SP_4$ being in functional group two and $SP_5$ and $SP_6$ being in functional group three. Each community was further replicated twice to account for the two treatments tested. The true underlying model was assumed to be the additive species interaction. The six identity effects were simulated from a $N(7,4)$ distribution while the six interaction terms were simulated from a $N(9,9)$ distribution with the contribution of the two treatment effects to the response being 5 and 8 respectively. The response was simulated for a subset of the original 50 simulation settings (10 $\theta$ and 5 $\sigma$ values) and 200 datasets were simulated for each simulation setting. The four different interaction structures from Table 1 in the original paper were fit to each of the datasets and the robustness of $\theta$ as well as the efficacy of the proposed model selection procedure were tested. Results are presented in Table C1(a) and Figure C1(a).

### b) 16 species model

The sixteen species design consisted of species $SP_1$, $SP_2$, $SP_3$,…, $SP_{16}$ spread across the simplex space in a design inspired from McKenna and Yurkonis 2016. The design consisted of total of 166 equi-proportional communities at richness levels of $1, 2, 4, 8,$ and $16$. Each community was replicated then twice, except for the 16 species mixture which was replicated 20 times. The functional grouping structure assumed was species $SP_1$, $SP_2$, $SP_3$, and $SP_4$ being in functional group one, $SP_5$, $SP_6$, $SP_7$, and $SP_8$ being in functional group two, $SP_9$, $SP_{10}$, $SP_{11}$ and $SP_{12}$ being in functional group three and $SP_{13}$, $SP_{14}$, $SP_{15}$, and $SP_{16}$ being in functional group four. The true underlying model was assumed to be the functional group model and the interaction effects were simulated keeping this functional grouping in mind, with the between functional group coefficients simulated from a $N(11,4)$ distribution and the within functional group coefficients simulated from a $N(5,4)$ distribution. The 16 identity effects were simulated from a $N(9,4)$ distribution. The response was simulated for a subset of the original 50 simulation settings (10 $\theta$ and 5 $\sigma$ values) and 200 datasets were simulated for each simulation setting. Four different functional group interaction models were fit to each of the 200 datasets for each simulation setting with a different functional grouping for each model. The four functional grouping tested were 'FG1' where all the species belonged to the same functional group, 'FG2' where there were two functional groups with species $SP_1, SP_2, ... SP_8$ and $SP_9, SP_{10}, ... SP_{16}$ being in the first and second functional grouping respectively, 'FG4' where four functional groups consisted of four species each with the same structure as the true underlying functional grouping and 'FG8' where there were eight functional groups, each consisting of two species. The goals of this study were to assess robustness of $\theta$ across the different functional groupings tested as well as the efficacy of the proposed new model section procedure. Results are presented in Table C1(b) and Figure C1(b).

c) **72 species model**

The 72 species model consisted of species $SP_1, SP_2, SP_3, ..., SP_{72}$ spread across the simplex space in a design consisting of 683 different communities ranging in richness from 1 up to 72. The design opted was the same used in Bell et al. 2005. Each community was repeated twice except for the 72 species mixture which was replicated 10 times. The functional grouping structure assumed species $SP_1, SP_2, ..., SP_{12}$ being in functional group one, $SP_{13}, SP_{14}, ..., SP_{24}$ being in functional group two, $SP_{25}, SP_{26}, ..., SP_{36}$ being in functional group three, $SP_{37}, SP_{38}, ..., SP_{48}$ being in functional group four, $SP_{49}, SP_{50}, ..., SP_{60}$ being in functional group five, and $SP_{61}, SP_{62}, ..., SP_{72}$ being in functional group six. The true underlying model was assumed to be the average pairwise interaction. The 72 identity effects were simulated from a $N(8, 4)$ distribution while the average interaction term was simulated from a $N(11, 4)$ distribution. The response was simulated for a subset of the original 50 simulation settings (10 $\theta$ and 5 $\sigma$ values) and 200 datasets were simulated for each simulation setting. The four different interaction structures from Table 1 in the original paper were fit to each of the datasets and the robustness of $\theta$ as well as the efficacy of the proposed model selection procedure were tested results of which are presented in Table C1(c) and in Figure C1(c).

Table C1. Simulation study results: The mean, standard deviation, coverage (conditional on convergence), and distribution of theta estimates across the 200 realizations for $\sigma = 1$ of each unique theta for each of the different models tested are presented. Similar results were observed for other $\sigma$ values too.

(a) Six species model: These estimates were generated with the true underlying model being the additive species model with identity effects 8, 6.8, 7.1, 5.7, 8, and 5.33 for species $SP_1, SP_2, SP_3, SP_4, SP_5$, and $SP_6$ respectively and interactions being simulated from normal distribution $N(9,9)$. The functional grouping structure assumed was $SP_1$ and $SP_2$ being in *FG1*, $SP_3$ and $SP_4$ being in *FG2* and $SP_5$ and $SP_6$ being in *FG3*. The four DI models (Average Pairwise, Functional Group, Additive Species and Full Pairwise) were fit to each dataset and results are presented.

| | Model | | | | | | | | | | | | | | | | |
|---|---|---|---|---|---|---|---|---|---|---|---|---|---|---|---|---|---|
| | Average Pairwise | | | | Functional Group | | | | Additive Species | | | | Full Pairwise | | | | |
| True Theta | Mean Est | SD Est | Coverage | Distribution | Mean Est | SD Est | Coverage | Distribution | Mean Est | SD Est | Coverage | Distribution | Mean Est | SD Est | Coverage | Distribution |
| 0.05 | 0.05 | 0.0024 | 0.95 | | 0.05 | 0.0025 | 0.955 | | 0.05 | 0.0024 | 0.96 | | 0.05 | 0.0024 | 0.96 | |
| 0.35 | 0.35 | 0.0057 | 0.945 | | 0.35 | 0.0058 | 0.945 | | 0.35 | 0.0057 | 0.95 | | 0.35 | 0.0058 | 0.945 | |
| 0.77 | 0.77 | 0.0152 | 0.945 | | 0.769 | 0.0153 | 0.945 | | 0.77 | 0.0152 | 0.96 | | 0.769 | 0.0153 | 0.96 | |
| 1 | 1 | 0.0253 | 0.955 | | 0.999 | 0.0253 | 0.955 | | 0.999 | 0.0253 | 0.96 | | 0.999 | 0.0253 | 0.955 | |
| 1.33 | 1.331 | 0.0593 | 0.945 | | 1.331 | 0.0599 | 0.94 | | 1.331 | 0.0588 | 0.95 | | 1.331 | 0.0592 | 0.96 | |

*(b)* Sixteen species model: These estimates were generated with the true underlying model being the functional group model with the 16 identity effects simulated from a $N(9,4)$ distribution for species $SP_1, SP_2, SP_3, \ldots, SP_{16}$ respectively. The functional grouping assumed was species $SP_1$, $SP_2$, $SP_3$, and $SP_4$ being in functional group one, $SP_5$, $SP_6$, $SP_7$, and $SP_8$ being in functional group two, $SP_9$, $SP_{10}$, $SP_{11}$ and $SP_{12}$ being in functional group three and $SP_{13}$, $SP_{14}$, $SP_{15}$, and $SP_{16}$ being in functional group four. The interaction coefficients were simulated by factoring this functional grouping with the between functional group coefficients simulated from a $N(11,4)$ distribution while the within functional group coefficients were simulated from a $N(5,4)$ distribution. The functional groups diversity interactions model was fit to each of the dataset with four different functional grouping, FG1 where all the species belonged to the same functional group, FG2 where there were two functional groups with species $SP_1, SP_2, \ldots SP_8$ and $SP_9, SP_{10}, \ldots SP_{16}$ being in the first and second functional grouping respectively, FG4 where four functional groups consisted of four species each with the same structure as the true underlying functional grouping and FG8 where there were eight functional groups, each consisting of two species.

| | Model | | | | | | | | | | | | | | | | |
|---|---|---|---|---|---|---|---|---|---|---|---|---|---|---|---|---|---|
| | FG1 | | | | FG2 | | | | FG4 | | | | FG8 | | | | |
| True Theta | Mean Est | SD Est | Coverage | Distribution | Mean Est | SD Est | Coverage | Distribution | Mean Est | SD Est | Coverage | Distribution | Mean Est | SD Est | Coverage | Distribution | |
| 0.05 | 0.05 | 0.0006 | 1 | | 0.053 | 0.0006 | 0.98 | | 0.05 | 0.0006 | 0.169 | | 0.05 | 0.0008 | 0.217 | | |
| 0.35 | 0.347 | 0.0021 | 0.988 | | 0.353 | 0.0022 | 0.912 | | 0.35 | 0.0023 | 0.442 | | 0.35 | 0.0024 | 0.428 | | |
| 0.77 | 0.761 | 0.0105 | 0.915 | | 0.772 | 0.0105 | 0.952 | | 0.77 | 0.011 | 0.797 | | 0.769 | 0.012 | 0.852 | | |
| 1 | 0.988 | 0.0169 | 0.872 | | 0.995 | 0.0164 | 0.918 | | 0.999 | 0.017 | 0.878 | | 0.998 | 0.0217 | 0.855 | | |
| 1.33 | 1.312 | 0.0431 | 0.947 | | 1.311 | 0.0421 | 0.932 | | 1.332 | 0.0456 | 0.965 | | 1.342 | 0.0875 | 0.914 | | |

(c) Seventy-two species model: These estimates were generated with the true underlying model being the average pairwise model with the 72 identity effects simulated from $N(8, 4)$ for species $SP_1, SP_2, SP_3, \ldots, SP_{72}$ respectively and the average interaction simulated from normal distribution $N(11,4)$. Six functional groups were assumed with species $SP_1, SP_2, \ldots, SP_{12}$ being in functional group one, $SP_{13}, SP_{14}, \ldots, SP_{24}$ being in functional group two, $SP_{25}, SP_{26}, \ldots, SP_{36}$ being in functional group three, $SP_{37}, SP_{38}, \ldots, SP_{48}$ being in functional group four, $SP_{49}, SP_{50}, \ldots, SP_{60}$ being in functional group five, and $SP_{61}, SP_{62}, \ldots, SP_{72}$ being in functional group six. The three DI models average pairwise, functional group, additive species were fit to each dataset and results are presented. The full pairwise model couldn't be fit to the data as there weren't enough communities in the experiment.

| | Model | | | | | | | | | | | | |
|---|---|---|---|---|---|---|---|---|---|---|---|---|---|
| | Average Pairwise | | | | Functional | | | | Additive | | | | |
| True Theta | Mean Est | SD Est | Coverage | Distribution | Mean Est | SD Est | Coverage | Distribution | Mean Est | SD Est | Coverage | Distribution | |
| 0.05 | 0.05 | 0.00003 | 0 | | 0.05 | 0.00003 | 0 | | 0.05 | 0.00003 | 0 | | |
| 0.35 | 0.35 | 0.00019 | 0 | | 0.35 | 0.0002 | 0 | | 0.35 | 0.00019 | 0 | | |
| 0.77 | 0.77 | 0.00218 | 0.44 | | 0.77 | 0.00219 | 0.435 | | 0.77 | 0.00219 | 0.41 | | |
| 1 | 1 | 0.00564 | 0.64 | | 1 | 0.00576 | 0.615 | | 1 | 0.00566 | 0.625 | | |
| 1.33 | 1.332 | 0.02108 | 0.885 | | 1.333 | 0.022 | 0.9 | | 1.333 | 0.02136 | 0.88 | | |

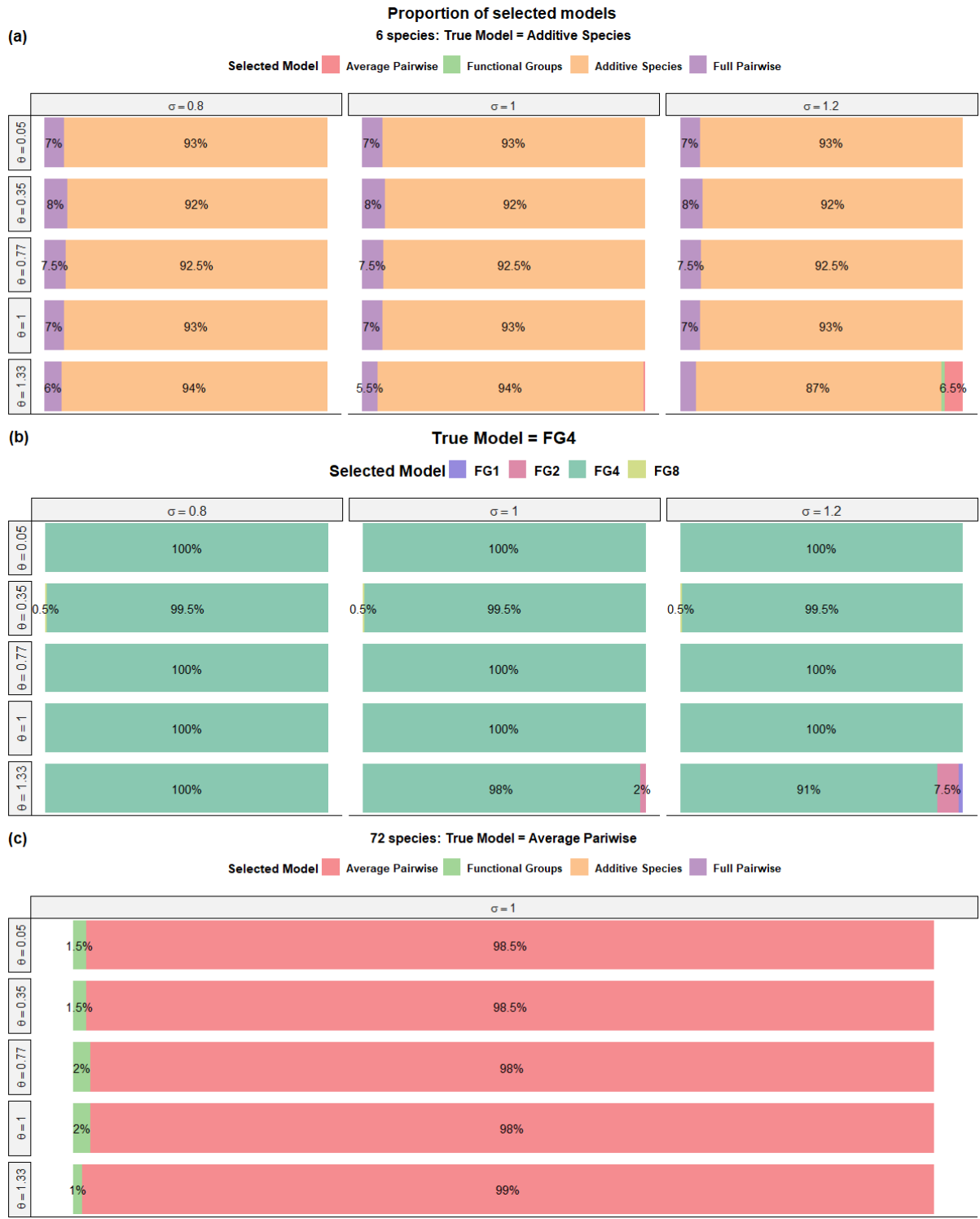

Figure C1. Efficacy of model selection procedure 'b': (a) For the six species model, (b) for the sixteen species model and (c) for the seventy-two species model. The proportion of times the different GDI models are selected across the 200 simulations for each unique $\theta - \sigma$ combination, when we first estimating $\theta$ for the simplest interaction structure and use the $\theta$ estimate from that model to fit and estimate the other models with interaction structures and then select the best model using AIC as the selection metric. The true underlying interaction structure for each model is highlighted above it.

## Situations when $\theta$ estimation can fail

The efficacy of the model selection procedure is primarily dependent on $\theta$ being robust across the different interaction structures. If the estimate of $\theta$ is significantly different across the different interaction structures, then the model selection procedure may fail. This could potentially happen if the underlying interaction terms themselves aren't significant or if there is a lack of fit in the model. These could have trickle down effects on the estimation of $\theta$ which could then affect the model selection. Further if the species communities aren't spread across the simplex space adequately, there wouldn't be enough information for the model, resulting in an incorrect estimate or even a failure to estimate $\theta$ for certain interaction structures. Table 1 shows examples of these issues. However, if we ensure that these issues aren't present in the data, then there shouldn't be any problem in estimating $\theta$ and selecting the appropriate interaction structure.

Table 1. Example scenarios where $\theta$ estimation and model selection can fail: Three different datasets were used and six different models were fit to the datasets including the identities only, average pairwise, functional groups, additive species, full pairwise diversity interactions models and the community factor model where each community is considered as a factor and a linear model is fit to the response, which is then used for the lack-of-fit test. The $\theta$ estimate and AIC for each of model is then presented along with the true underlying (marked by *) and the selected model (shaded in grey) highlighted in the table.

(a) Non-significant interaction terms: The `sim1` dataset from the `DImodels` package which is a four species experiment having no significant interaction effects is used and the aforementioned six models are fit.

| Model | θ Estimate | AIC |
|---|---|---|
| Identities only* | NA | 206.14 |
| Average Pairwise | 1.5 | 207.85 |
| Functional Groups | 1.5 | 213.06 |
| Additive Species | 1.5 | 211.68 |
| Full Pairwise | 1.5 | 215.67 |
| Community Factor | NA | 213.58 |

*True underlying model*
*Shaded model is the selected model*

(b) Lack of fit in models: The same design as the `sim1` dataset is used but the response is simulated by assigning a different mean to each community whilst ignoring the species proportions. Thus, there aren't any species interactions that are getting factored into the response. The six models are then fit to the data and the results are presented.

| Model | θ Estimate | AIC |
|---|---|---|
| Identities only | NA | 658.07 |
| Average Pairwise | 1.5 | 658.72 |
| Functional Groups | 1.5 | 658.21 |
| Additive Species | 1.5 | 642.36 |
| Full Pairwise | 1.5 | 643.72 |
| Community Factor* | NA | 461.40 |

*True underlying model*

(c) Inadequate coverage across the simplex: One specific example simulation setting from the 1000 simulation settings for the four species simulation in the original paper is taken. It is assumed that the true value of $\theta$ is 0.91 with the true underlying interaction structure being the full pairwise model. However instead of taking the entire design from the original simulation a subset of the design that includes only the monocultures, and the four species mixtures is taken and the six models are then fit to the data.

| Model | θ Estimate | AIC |
|---|---|---|
| Identities only | NA | 338.60 |
| Average Pairwise | 0.6366 | 264.58 |
| Functional Groups | 0.6018 | 268.00 |
| Additive Species | 0.5862 | 265.82 |
| Full Pairwise* | 0.5862 | 268.56 |
| Community Factor | NA | 282.89 |

*\* True underlying model*
*Shaded model is the selected model*